\documentclass[floatfix,amssymb,prl,twocolumn,superscriptaddress,nofootinbib, aps]{revtex4-2}	
\usepackage{amssymb,amsmath,amsthm,mathrsfs,amsfonts,pifont,verbatim,mathtools,needspace,enumitem,etoolbox,graphicx,physics,microtype,afterpage,bm,soul}
\usepackage[dvipsnames]{xcolor}
\definecolor{linkcolor}{rgb}{0.0,0.3,0.5}
\usepackage[unicode, colorlinks=true, linkcolor=linkcolor, citecolor=linkcolor, filecolor=linkcolor,urlcolor=linkcolor, pdfusetitle]{hyperref}
\usepackage{orcidlink}
\usepackage[all]{hypcap}
\usepackage[T1]{fontenc}
\usepackage[utf8]{inputenc}
\usepackage{tabularx}
\usepackage{tensor}
\usepackage{tikz-cd}
\usepackage{cancel}
\newcommand{\tho}{\text{\th}}
\renewcommand{\eth}{\text{\dh}}
\newcommand{\thop}{\text{\th}^\prime}

\usepackage{lmodern}
\usepackage{ragged2e}
\allowdisplaybreaks
\usepackage{tikz}
\usepackage{color}
\usepackage{framed}
\usepackage{hyperref}
\hypersetup{colorlinks, citecolor=darkblue, linkcolor=black, urlcolor=darkblue}
\definecolor{rossos}{cmyk}{0,1,1,0.55}
\definecolor{bluscuro}{rgb}{0.15, 0.2, .85}
\definecolor{bluchiaro}{cmyk}{1,.3,0.,0.1}
\definecolor{ForestGreen}{rgb}{0.13, 0.55, 0.13}
\definecolor{darkblue}{rgb}{0,0, 1.39}

\newcommand{\be}{\begin{equation}}
\newcommand{\ee}{\end{equation}}

\def\lsim{\mathrel{\rlap{\lower4pt\hbox{\hskip0.5pt$\sim$}}
    \raise1pt\hbox{$<$}}}         
\def\gsim{\mathrel{\rlap{\lower4pt\hbox{\hskip0.5pt$\sim$}}
    \raise1pt\hbox{$>$}}}         

\newcommand{\jhu}{William H.\ Miller III Department of Physics and Astronomy, Johns Hopkins University, \\ 3400 North Charles Street, Baltimore, Maryland, 21218, USA}

\begin{document}

\title{Gravitational electric-magnetic duality at the light ring \\ and quasinormal mode isospectrality in effective field theories}

\author{Ibrahima Bah\orcidlink{0000-0003-2843-4401}}
\email{iboubah@jhu.edu}
\affiliation{\jhu}

\author{Emanuele Berti\orcidlink{0000-0003-0751-5130}}
\email{berti@jhu.edu}
\affiliation{\jhu}

\author{Valerio De Luca\orcidlink{0000-0002-1444-5372}}
\email{vdeluca2@jh.edu}
\affiliation{\jhu}

\author{Bogdan Ganchev\orcidlink{0000-0001-7081-8156}}
\email{bganche1@jh.edu}
\affiliation{\jhu}

\author{David Pereñiguez\orcidlink{0000-0002-1007-4551}}
\email{dpereni1@jh.edu}
\affiliation{\jhu}


\begin{abstract}
\medskip
\noindent
Black hole perturbations are characterized by a superposition of damped exponentials known as quasinormal modes. In general relativity, the spectra of parity-even and parity-odd quasinormal modes coincide --- a property known as isospectrality, which is typically broken by corrections beyond general relativity. Recently, certain higher-derivative operators were shown to preserve isospectrality in the high-frequency (eikonal) regime. Motivated by the relation between the light ring Penrose limit and the eikonal limit, we study isospectrality in a class of plane-wave spacetimes. In general relativity, we show that dynamical metric fluctuations on these backgrounds admit a gravitational analog of electric-magnetic duality, which enforces isospectrality. Requiring this duality to persist in the presence of higher-derivative corrections constrains the couplings so that isospectrality is  preserved. 
We conclude that gravitational electric-magnetic duality at the light ring is the organizing principle behind isospectrality in the eikonal limit, and we conjecture that this remains true for other duality-invariant corrections to general relativity.

\end{abstract}
\maketitle

\vspace{0.1cm}
\noindent{{\bf{Introduction.}}}
Gravitational-wave astronomy has opened an unprecedented window onto the strong-field dynamics of spacetime. The LIGO-Virgo-KAGRA Collaboration runs have yielded an expanding catalog of compact-binary merger observations in which the short, clean ringdown stage is becoming increasingly amenable to precision measurements~\cite{Berti:2025hly}. This stage is dominated by a discrete set of damped oscillations known as quasinormal modes (QNMs)~\cite{Vishveshwara:1970zz,Press:1971wr,Teukolsky:1973ha,Chandrasekhar:1975zza,Kokkotas:1999bd,Berti:2009kk}. Because QNM frequencies are fixed entirely by the background spacetime and by the field equations that govern small perturbations, they provide a direct and model-independent probe of gravity in its most nonlinear regime~\cite{Dreyer:2003bv,Berti:2005ys,Gossan:2011ha, Berti:2015itd,Berti:2018vdi,Franchini:2023eda,LIGOScientific:2026qni,LIGOScientific:2026wpt}.

A remarkable structural feature of QNM spectra in general relativity (GR) is isospectrality between perturbation sectors of different parity. In spherically symmetric black holes, parity-even and parity-odd gravitational perturbations are governed by different master equations, the Regge-Wheeler~\cite{Regge:1957td} and Zerilli~\cite{Zerilli:1970se} equations, which both take a Schrödinger-like form and share the same spectrum. Chandrasekhar made this equivalence explicit by constructing transformations that map the solutions of one equation into solutions of the other~\cite{Chandrasekhar:1975nkd}. Subsequent work showed that these mappings can be understood as Darboux (or supersymmetric quantum-mechanical) transformations, linking isospectrality to integrability-type structures~\cite{Kallosh:1997ug,Glampedakis:2017rar,Lenzi:2021njy,Lenzi:2021wpc,Lenzi:2022wjv,Lenzi:2023inn,Solomon:2023ltn}. Extensions of these ideas to the rotating (Kerr) case, using generalized Darboux transformations, have further clarified how different perturbation channels are related in more general spacetimes~\cite{Nichols:2012jn, Glampedakis:2017rar, Franchini:2023xhd}. Alternatively, QNM isospectrality can be seen as a consequence of the dynamics being governed by self-dual curvature variables (see~\cite{Mukkamala:2024dxf}, which is closer in spirit to the perspective of the present work). 

However, isospectrality is fragile. In general, effective field theories (EFTs) of gravity --- which incorporate higher-curvature corrections encoding short-distance effects --- alter the linearized equations of motion and can break isospectrality~\cite{Cardoso:2019mqo, McManus:2019ulj, Cano:2021myl, Cano:2023jbk, Silva:2024ffz, Cano:2024ezp, Silva:2026jih}. Recent computations of high-frequency (eikonal) QNMs in broad classes of EFTs have shown that isospectrality is preserved only for special combinations of higher-derivative operators, while generic corrections produce different spectra in the eikonal limit~\cite{Cano:2024wzo, Cano:2025mht}. Isospectrality-preserving theories turn out to have the form predicted by type II string theory curvature corrections. 
This raises two natural questions of theoretical and observational relevance: what geometric or symmetry principle enforces isospectrality in the eikonal regime, and which beyond-GR deformations preserve or break it?

We argue that \textit{gravitational} electric-magnetic duality provides a useful conceptual perspective to address this question. In Maxwell theory, the source-free equations and energy-momentum tensor are invariant under rotations that mix electric and magnetic fields --- equivalently, rotations of the field strength into its Hodge dual~\cite{Gaillard:1981rj, Gibbons:1995cv}. In quantum field theory and string theory, the broader notion of S-duality relates weak and strong coupling regimes and exchanges electric and magnetic objects, and it yields stringent constraints on spectra and interactions~\cite{Kol:2023yxd}. When the coupling constants are fixed, the duality reduces to a symmetry of the theory, so in this paper we will refer to such dualities and to the corresponding symmetries interchangeably. In gravity, one can formulate an analogous duality by rotating the Riemann tensor into its Hodge dual at the perturbative level; when that is possible, it can strongly constrain both background solutions and their linearized perturbations~\cite{Penrose:1960eq,Nieto:1999pn,Hull:2001iu,Henneaux:2004jw,Deser:2005sz,Bunster:2006rt,Argurio:2009xr,Barnich:2008ts,Boos:2021suz,Kol:2023yxd, Monteiro:2023dev,delRio:2025ynh}.

Here we connect these ideas to eikonal QNMs using two complementary ingredients. The first is geometric: in GR and in the eikonal limit, QNMs are localized on unstable null geodesics (light rings or photon spheres) and their frequencies are controlled by the orbital frequency and Lyapunov exponent of these null trajectories~\cite{1972ApJ...172L..95G, Ferrari:1984zz, Mashhoon:1985cya, Berti:2005eb, Cardoso:2008bp, Dolan:2010wr}. The Penrose limit provides the precise tool to capture this local light ring geometry: zooming in on a null geodesic produces a universal plane-wave (pp-wave) spacetime that isolates the environment responsible for eikonal dynamics~\cite{Penrose1976, Fransen:2023eqj, Kapec:2024lnr}. The second ingredient is symmetry: on these plane waves the Riemann tensor admits a natural (anti-)self-dual decomposition, and linearized metric perturbations are encoded in a single, complex Hertz potential $\Psi_\text{\tiny H}$~\cite{Fransen:2025cgv}. We show that a global phase rotation of $\Psi_\text{\tiny H}$ implements the linearized analog of S-duality --- it rotates (anti-)self-dual curvature components into one another --- and that this phase symmetry forces 
different parity sectors to obey the same linear dynamics and boundary conditions, giving rise to isospectrality. 

Building on this observation, we show that S-duality on the light ring provides a simple symmetry criterion for EFT deformations: only corrections that respect the gravitational duality avoid coupling the Hertz potential to its complex conjugate and thus preserve the eikonal parity degeneracy, while generic duality-breaking operators induce such couplings and split the spectra. This symmetry perspective not only explains recent empirical findings (including special eight-derivative combinations that retain isospectrality~\cite{Cano:2024wzo}), but also gives a practical diagnostic to judge which beyond-GR modifications will affect high-frequency ringdown signals.

The paper is organized as follows. First, we review the eikonal connection between QNMs and unstable null geodesics and introduce the Penrose limit that yields the local pp-wave geometry near a light ring. Next, we present the Hertz-potential formulation of linearized metric perturbations on pp-waves, show that its global phase rotation realizes the linearized gravitational S-duality, and explain how this symmetry enforces isospectrality 
in the eikonal approximation. Building on this result, we translate the duality requirement into constraints that higher-derivative EFT corrections must satisfy to preserve isospectrality. 
We adopt geometric units $G=c=1$ and the mostly minus convention $(+,-,-,-)$, and we define the Riemann's sign so that $[\nabla_{a},\,\nabla_{b}]X_{c}=R_{abcd}X^{d}$.

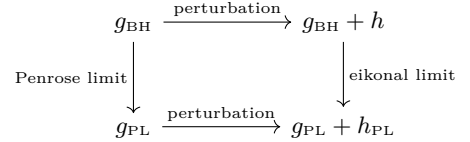
\begin{figure}[ht]
\centering
\begin{tikzcd}[column sep=huge, row sep=large]
g_\text{\tiny BH} \arrow[r, "\text{perturbation}"] \arrow[d, "\text{Penrose limit}"'] 
& g_\text{\tiny BH}+h \arrow[d, "\text{eikonal limit}"] \\
g_\text{\tiny PL} \arrow[r, "\text{perturbation}"] 
& g_\text{\tiny PL}+h_\text{\tiny PL}
\end{tikzcd}
\caption{Relation between black hole and Penrose-limit perturbations (see~\cite{Fransen:2023eqj}).}
\label{fig:bh-pl-diagram}
\end{figure}

\vspace{0.1cm}
\noindent{{\bf{The eikonal regime and Penrose-limit pp-waves.}}}
The high-frequency (eikonal) QNMs of a black hole are determined by waves localized near unstable null orbits (the light ring); consequently, the local geometry along a light ring geodesic determines the leading eikonal spectrum. The Penrose limit of a spacetime $(M,g_{ab})$ along a null geodesic $\gamma(u)$ (e.g., a black hole light ring geodesic) captures precisely this local geometry and yields the pp-wave~\cite{Penrose1976, Fransen:2023eqj, Kapec:2024lnr}
(in Brinkmann coordinates~\cite{Brinkmann:1925fr})
\begin{equation}\label{eq:PL}
{\rm d}s^{2}=-A_{ij}(u)x^{i}x^{j}{\rm d}u^{2}+2{\rm d}u{\rm d}v-{\rm d}x^{2}-{\rm d}y^{2}\, .
\end{equation}
Here $x^{i}=(x,y)$, and the $u$-dependent matrix $A_{ij}(u)$ is obtained from the Riemann tensor of the original spacetime, $\text{R}_{abcd}$, via $ A_{ij}(u)=\text{R}_{abcd}E^{a}_{i}E^{b}_{+}E^{c}_{j}E^{d}_{+}\rvert_{\gamma(u)}$, where $(E^{a}_{+},\,E^{a}_{-},\,E^{a}_{i})$ is a null frame of the original spacetime, with $E^{a}_{+}$ tangent to a null congruence containing $\gamma$ and $E^{a}_{-},\,E^{a}_{i}$ parallelly propagated along $E_{+}$. Penrose showed that the geometry~\eqref{eq:PL} follows after a well-defined ``zoom in'' procedure on the spacetime in the vicinity of $\gamma(u)$~\cite{Penrose1976}.
Fransen suggested using this observation operationally: first take the Penrose limit along the light ring, then study the characteristic modes of the resulting pp-wave~\cite{Fransen:2023eqj}. The claim is that this procedure reproduces the eikonal QNMs of the full black hole geometry, or (more formally) that the diagram in Fig.~\ref{fig:bh-pl-diagram} commutes.

This correspondence has been checked at the linear~\cite{Fransen:2023eqj} and nonlinear~\cite{Fransen:2025cgv,Kehagias:2025ntm,Perrone:2025zhy} level for QNMs in GR, and has been extended also to horizon physics~\cite{Kapec:2024lnr}. Motivated by this link to the eikonal regime, we analyze isospectrality in the class of spacetimes given by Eq.~\eqref{eq:PL}, first in GR and then in theories with higher-curvature corrections.

It is convenient to work in a null tetrad adapted to the pp-wave, and to use the Geroch-Held-Penrose (GHP) formalism~\cite{Geroch:1973am} (see also~\cite{Stewart:1974uz, Green:2022htq, Aksteiner:2010rh, Aksteiner:2016pjt}). The GHP framework packages directional derivatives and spin coefficients in a frame-covariant way and is particularly useful when the background admits a preferred null congruence. For the Brinkmann background of Eq.~\eqref{eq:PL}, a convenient null frame is (see the Appendix of Ref.~\cite{Fransen:2025cgv} for a summary)
\begin{equation}\label{eq:nullFrame}
    \ell=\partial_{v}\,,~ n=\partial_{u}+\frac{1}{2}A_{ij}x^{i}x^{j}\partial_{v}\, ,~ m=\frac{1}{\sqrt{2}}\left(\partial_{x}+i\partial_{y}\right)\,,
\end{equation}
where the null vector $\ell$ is parallel propagated along itself, and the remaining frame vectors are chosen to be parallel propagated along it ($\nabla_{\ell} \ell=\nabla_{\ell} n=\nabla_{\ell} m=0$).
$\ell$ aligns with the repeated principal null direction of \eqref{eq:PL}, which is Petrov type-$N$, hence all Weyl scalars $\Psi_{i}$ vanish except $\Psi_{4}$. Finally, the geometry is Ricci-flat when
\begin{equation}\label{eq:OnShell}
A_{22}(u)=-A_{11}(u)\equiv A(u)\,.
\end{equation}
More details are given in the Supplemental Material (SM).
In the following, we study the symmetry structure of pp-wave spacetimes and the appearance of the electric-magnetic duality.

\vspace{0.1cm}
\noindent{{\bf{Electric-magnetic duality in gravity.}}} 
We begin by recalling the familiar structure of the electric-magnetic duality in Einstein-Maxwell theory, phrased so as to make the analogy with gravity manifest. In this theory, electric-magnetic duality is the statement that given a solution to the field equation with field strength $F_{ab}$, Hodge dual $\star F_{ab}=\frac{1}{2}\epsilon_{abcd}F^{cd}$, and associated (anti-)self-dual $F^{\pm}=\left(F\mp i\star F\right)/2$, 
a new solution emerges upon the transformation
\begin{equation}\label{eq:F_duality}
    F^{\pm}\mapsto e^{\pm i\theta}F^{\pm}\, .
\end{equation}
This symmetry is a nonperturbative, on-shell statement: it follows from the linear (vacuum) Maxwell equations  $dF^{\pm}=0$ and the invariance of the energy-momentum tensor $T\sim F^{+}F^{-}$, and it corresponds to the SO(2) subgroup of the S-duality group SL$(2,\mathbb{R})$ associated with the theory.
This U(1) duality enforces helicity conservation in photon scattering amplitudes~\cite{1965AmJPh..33..958C}.

In some cases, this can be extended to the gravitational field (see e.g.~\cite{Monteiro:2023dev}). The idea is to promote the (anti-)self-dual Weyl tensor $C^{\pm}=\left(C\mp i\star C\right)/2$, which
encodes opposite graviton helicities, to the role of a field strength, and assess whether the analog of Eq.~\eqref{eq:F_duality}, 
\begin{equation}\label{eq:C_duality}
    C^{\pm}\mapsto e^{\pm i\theta}C^{\pm}\, ,
\end{equation}
is a symmetry of the equations of motion. As already noted by Penrose~\cite{Penrose:1960eq} (see also the discussion in Ref.~\cite{Monteiro:2023dev}), it is clear that this cannot be achieved exactly in vacuum GR, where $C^{\pm}$ satisfies an equation of the form $\square C^{\pm}+(C^{\pm})^{2}=0$, where the self-interacting term $(C^{\pm})^{2}$ transforms differently under Eq.~\eqref{eq:C_duality} than $\square C^{\pm}$ (in contrast to the Einstein--Maxwell case).

There are, however, a few notable exceptions, which fall into two categories. In a certain subset of exact solutions the duality can hold nonperturbatively. For example, the duality rotations~\eqref{eq:C_duality} in the Kerr--Taub--NUT family correspond to rotating the mass and NUT charge of the solution. This symmetry is not generic, though: small perturbations typically break it. By contrast, in other cases the duality holds only perturbatively at the level of linear fluctuations, but it is otherwise universal in that every fluctuation respects the symmetry.  Examples include fluctuations about flat spacetime~\cite{Henneaux:2004jw,delRio:2025ynh}, where the quadratic Einstein-Hilbert action for the two physical helicity modes $h^+$ and $h^-$, in the light-cone gauge, takes the chiral form~\cite{Ananth:2006fh}
\begin{equation}
\label{eq.actionMonteiro}
S_\text{\tiny GR,flat}[h^+,h^-]=\int {\rm d}^4x\;h^-\,\Box\,h^+ + \mathcal{O}(h^3)\,.
\end{equation}
This is manifestly invariant under the phase rotation $h^\pm \mapsto e^{\pm i\theta} h^\pm$ associated with Eq.~\eqref{eq:C_duality}, although only at linear level --- GR nonlinearities break duality invariance.

In this work we show that gravitational duality holds in both senses described above for Penrose-limit pp-waves \eqref{eq:PL}. At the level of the background, a straightforward computation shows that a left SO(2) rotation, $\mathcal{R}_{ik}(\theta)$, acting on the transverse matrix via $A_{ij}(u)\mapsto\mathcal{R}_{ik}(\theta)\,A_{kj}(u)$ induces the duality transformation of the (anti-)self-dual Weyl tensor given in \eqref{eq:C_duality} if and only if the Ricci-flatness condition \eqref{eq:OnShell} is satisfied. Note that this is not just a coordinate rotation of $x^{i}$, which would instead yield $A\mapsto\mathcal{R}A\mathcal{R}^T$. More importantly, and technically harder, is proving that dynamical linear fluctuations around the background \eqref{eq:PL} are also duality invariant. We address this next and show that the statement holds in GR and in a specific class of higher-derivative theories compatible with Type II string-theory corrections~\cite{Schwarz:1995dk}.

\vspace{0.1cm}
\noindent{{\bf{Electric-magnetic duality for fluctuations of Penrose limits.}}} 
For type-$N$ backgrounds such as our pp-wave \eqref{eq:PL}, it is convenient to describe linearized metric perturbations using the 2-spinor and GHP formalisms. In this setting, vacuum solutions for the metric perturbation $h_{ab}$ are encoded in a single complex Hertz potential $\Psi_\text{\tiny H}$ carrying GHP weight $(-4,0)$ (analogous to one of the Hertz potentials of type-$D$ spacetimes \cite{Wald:1978vm,Kegeles}). The metric perturbation is then recovered from $\Psi_\text{\tiny H}$ via the reconstruction map~\cite{Fransen:2025cgv}
\begin{align}
\label{eq:recmet2_final}
h_{ab}={}& -\ell_a\ell_b\big(\eth^2\Psi_\text{\tiny H}+\eth'^2\bar\Psi_\text{\tiny H}\big)  -m_a m_b\,\tho^2\Psi_\text{\tiny H} \\
& -\bar m_a\bar m_b\,\tho^2\bar\Psi_\text{\tiny H}+2\ell_{(a}m_{b)}\tho\eth\Psi_\text{\tiny H}+2\ell_{(a}\bar m_{b)}\tho\eth'\bar\Psi_\text{\tiny H} \nonumber \,.
\end{align}
It satisfies the transverse, traceless and Lorenz conditions ($\ell^a h_{ab}=0$,\,$\nabla^a h_{ab}=0$,\,$h^a{}_a=0$).
The equation of motion for the Hertz potential is simply $\square \Psi_\text{\tiny H}=0$ \cite{Fransen:2025cgv}. To compare with \eqref{eq.actionMonteiro}, we can expand the Einstein-Hilbert action to quadratic order in $\Psi_\text{\tiny H}$ on the background~\eqref{eq:PL}, without using the equation of motion of $\Psi_\text{\tiny H}$, to get
\begin{equation}
\begin{aligned}\label{eq:secondorderaction_final}
S_\text{\tiny GR,pp}&=-\frac{3}{2\pi}\int {\rm d}^{4}x\sqrt{|g|} \Psi_\text{\tiny H} \,\Box \,\tho^{4}\bar \Psi_\text{\tiny H}\,.
\end{aligned}
\end{equation}
This is analogous to the gravitational action for perturbations of flat space in Eq.~\eqref{eq.actionMonteiro}, but now generalized to perturbations about the metric \eqref{eq:PL}. It is manifest that the transformation 
\begin{equation}\label{eq:EMDualityHertz_final}
\Psi_\text{\tiny H}\mapsto e^{\pm i\theta}\Psi_\text{\tiny H}
\end{equation}
is a symmetry of the  equations of motion. In fact, such a symmetry in the Hertz potential is also present for perturbations of Kerr black holes. However, as we show next, in the background \eqref{eq:PL} this symmetry also corresponds to a gravitational electric-magnetic duality transformation --- unlike in Kerr, and in analogy with the flat space case. To see this, we notice that the perturbations of the curvature Weyl scalars are given in terms of $\Psi_\text{\tiny H}$ via
\begin{equation}\label{eq:Weylperts}
    \dot{\Psi}_{n}=-\frac{1}{2}\tho^{(4-n)}\eth'^{n}\bar{\Psi}_\text{\tiny H}\, ,\quad (n=0,1,2,3,4)\, ,
\end{equation}
where in this paper an overdot denotes a linear perturbation (except for $\dot{g}_{ab}$, where we use the customary $h_{ab}$ instead). Hence, they transform as $ \dot{\Psi}_{n}\mapsto e^{-i \theta }\dot{\Psi}_{n}$ under \eqref{eq:EMDualityHertz_final}. As shown in \cite{Fransen:2025cgv}, one also has that $\dot{m}_{a}\mapsto e^{-i\theta}\dot{m}_{a}$ and from the expression of $\dot{C}^{\pm}_{abcd}$ in terms of Weyl scalars and frames (see SM for the explicit expression), one finds
\begin{equation}\label{eq:WeylTrafo}
    \dot{C}^{\pm}_{abcd}\mapsto e^{\pm i\theta}\dot{C}^{\pm}_{abcd}\, .
\end{equation}
This shows that \textit{gravitational} electric-magnetic duality is an on-shell symmetry of  dynamical perturbations of Penrose limit pp-waves. Given that this framework captures the physics of black hole fluctuations in the eikonal limit, we uncover gravitational duality as a key property in that regime. Next, we explore the consequences of this duality for QNM isospectrality, both in GR and its EFT corrections. 

\vspace{0.1cm}

\noindent{{\bf{Eikonal isospectrality from electric-magnetic duality in GR.}}} 
The symmetry \eqref{eq:EMDualityHertz_final} is nontrivial: it is not the mere scaling symmetry of free linear fluctuations, but a genuine U(1) rotation of the complex phase. Any wave operator that mixed $\Psi_\text{\tiny H}$ and its complex conjugate would violate this invariance, so the absence of such mixing terms in \eqref{eq:secondorderaction_final} is significant. Here, we show that it is responsible for QNM isospectrality in the class of pp-waves within \eqref{eq:PL}, that describe the eikonal regime of black hole QNMs. These are the subclass where $A_{12}(u)=0$, $A_{22}(u)=-A_{11}(u)\equiv\Omega^{2}$, and $\Omega$ is a constant related to the light ring frequency \cite{Fransen:2023eqj}:
\begin{equation}\label{eq:PLLR}
    {\rm d}s^{2}=\Omega^{2}\left(x^{2}-y^{2}\right){\rm d}u^{2}+2{\rm d}u{\rm d}v-{\rm d}x^{2}-{\rm d}y^{2}\,.
\end{equation}
These geometries can be thought of as a gravitational version of a simple and inverted harmonic oscillator, where confining happens along $x$ and dispersion along $y$ (corresponding to the polar and radial directions of the original black hole spacetime). Indeed, the isometry algebra is generated by the (complex-valued) Killing vectors~\cite{Hadar:2022xag, Fransen:2023eqj,Fransen:2025cgv}
\begin{align}
a_{\pm}&=\frac{e^{\pm i \Omega u}}{\sqrt{2 \Omega}}\left(\pm i \Omega x \partial_{v}+\partial_{x}\right)\, ,~ b_{\pm}=\frac{e^{\mp \Omega u}}{\sqrt{2 \Omega}}\left(\mp \Omega y \partial_{v}+\partial_{y}\right)\, , \nonumber\\ \mathbf{1}&=i\partial_{v}\, , \quad H_{\pm}=\mp i \partial_{u}\, ,
\end{align}
which span the simple and inverted harmonic oscillator algebras (see also~\cite{Despontin:2025svw}). The QNM solutions come from a Hertz potential of the form~\cite{Fransen:2025cgv}
\begin{equation}\label{eq:Hertz_Sep}
    \Psi_\text{\tiny H}=e^{i(p_{v} v+p_{u} u)}X_{n_{x}}(x)Y(y)\, ,
\end{equation}
where $X_{n_{x}}(x)$ corresponds to a solution of the simple harmonic oscillator with energy level $n_{x}\in Z_{\geq0}$, and $p_{u}$ plays the role of the QNM frequency, as established in~\cite{Fransen:2023eqj,Fransen:2025cgv}. In the following, we want to understand what Hertz potentials \eqref{eq:Hertz_Sep} generate gravitational waves with definite parity with respect to the orientation-reversing isometry 
\begin{equation}\label{eq:Zmap}
    \mathcal{Z}_{2}\colon (u,v,x,y) \mapsto (u,v,-x,y)\, .
\end{equation}
This corresponds to the usual parity transformation $\theta\mapsto\pi-\theta$ in the original black hole spacetime. Define the parity operator as $P=(-1)^{n_{x}}\mathcal{Z}_{2*}$, where $\mathcal{Z}_{2*}$ is the pull-back of \eqref{eq:Zmap} (the factor $(-1)^{n_{x}}$ is included so that a Hertz potential of the form \eqref{eq:Hertz_Sep} satisfies $P\Psi_\text{\tiny H}=+\Psi_\text{\tiny H}$). Then, any metric perturbation can be decomposed into its parity-even and parity-odd sectors,
\begin{equation}
    h_{\mu\nu}=h^{+}_{\mu\nu}+h^{-}_{\mu\nu}\,,\quad h^{\pm}_{\mu\nu}\equiv \frac{1}{2} \left(h_{\mu\nu}\pm P h_{\mu\nu} \right)\,,
\end{equation}
and since $P$ commutes with the equations of motion, it is guaranteed that they should satisfy decoupled equations. Given a Hertz potential \eqref{eq:Hertz_Sep}, it is straightforward to show that the new potentials
\begin{equation}
    \Psi_\text{\tiny H}^{\pm}\equiv \frac{1}{2}\left(1\pm PC\right)\Psi_\text{\tiny H}\, ,
\end{equation}
where $C$ is complex conjugation, generate the even and odd parts of the metric perturbation $h^{\pm}\left(\Psi_\text{\tiny H}^{\pm}\right)$ through the map \eqref{eq:recmet2_final} (see SM for details). Separating $\Psi_\text{\tiny H}$ into its real and imaginary parts $\Psi_\text{\tiny H}=\Psi^{\text{\tiny R}}_\text{\tiny H}+i\Psi^{\text{\tiny I}}_\text{\tiny H}$, it is found that $\Psi_\text{\tiny H}^{+}=\Psi^{\text{\tiny R}}_\text{\tiny H}$ and $\Psi_\text{\tiny H}^{-}=i\Psi^{\text{\tiny I}}_\text{\tiny H}$, so we learn that even (odd) gravitational perturbations are generated by purely real (imaginary) Hertz potentials.  

Isospectrality is the statement that the real and imaginary Hertz modes $\Psi^\text{\tiny R}_\text{\tiny H}$ and $\Psi^\text{\tiny I}_\text{\tiny H}$ have the same eigenfrequencies. Both modes satisfy the same boundary conditions (regularity at $y=0$ and purely outgoing behavior as $|y|\to\infty$), so equality of their spectra follows if they obey the same wave equation. The symmetry \eqref{eq:EMDualityHertz_final} guarantees that this is true by forbidding the mixing of $\Psi_\text{\tiny H}$ and $\bar{\Psi}_\text{\tiny H}$, and hence even and odd sectors are isospectral. There is an analog argument for perturbations in Kerr, where isospectrality can be seen as a consequence of the symmetry operator of Eq.~\eqref{eq:EMDualityHertz_final}, see e.g.~\cite{Li:2023ulk}. 
However, it is in our light ring pp-wave framework \eqref{eq:PLLR}, capturing the eikonal regime of QNMs, that it admits an additional geometric interpretation: gravitational electric-magnetic duality invariance \eqref{eq:WeylTrafo}. For completeness, we recall that the QNM frequencies read~\cite{Fransen:2025cgv}
\begin{equation}\label{eq:onshell_freq}
p_u = \mathrm{sgn}(p_v) \, \Omega\left(n_x + \frac{1}{2}\right) + i \Omega \left(n_y + \frac{1}{2}\right) \,,
\end{equation}
where $n_{y}$ is a nonnegative integer (the overtone number). The precise dictionary between these QNMs and the eikonal Kerr ones is spelled out in~\cite{Fransen:2023eqj}.

\vspace{0.1cm}
\noindent{{\bf{Eikonal isospectrality from electric-magnetic duality in EFTs.}}}
Here, we consider higher-derivative operators extending GR and show that, if one requires that on-shell gravitational duality invariance is preserved, then only specific combinations of operators remain. These are precisely the operators that preserve isospectrality. Up to eight-derivative corrections, the most general EFT extension of GR (discarding parity-breaking operators for reasons explained below) reads~\cite{Endlich:2017tqa,Cano:2019ore, Cano:2023jbk}
\begin{equation}\label{eq:EFTAction}
\begin{aligned}
S_\text{\tiny EFT}&=\frac{1}{16\,\pi}\int {\rm d}^4x\,\sqrt{\lvert g\rvert}\Big[R+\lambda_{\rm ev}\,\mathcal{I}+\epsilon_1\,\mathcal{C}^2+\epsilon_2\,\tilde{\mathcal{C}}^2\Big]\,,
\end{aligned}
\end{equation}
with $\mathcal{I}=\tensor{R}{_{ab}^{cd}}\,\tensor{R}{_{cd}^{ef}}\,\tensor{R}{_{ef}^{ab}}$, $\mathcal{C}=R_{abcd}\,R^{abcd}$ and $\tilde{\mathcal{C}}=R_{abcd}\,\left(\star R\right)^{abcd}$. The dimensional couplings $\lambda_{\text{ev}}$ and $\epsilon_{1,2}$ scale with $\ell_\text{\tiny UV}^{4}$ and $\ell_\text{\tiny UV}^{6}$, where $\ell_\text{\tiny UV}$ is the length scale of new physics. 

First, we note that the geometries \eqref{eq:PL} are exact solutions of this theory, so they can be taken as genuine backgrounds. In general, though, dynamical fluctuations will feel the higher-derivative operators. We focus on the subclass of backgrounds \eqref{eq:PLLR} and study the effect of the corrections on isospectrality with respect to the parity isometry \eqref{eq:Zmap} (if we had included parity-breaking corrections, then $P$ would not commute with the equations of motion, yielding coupled even and odd gravitational sectors; no notion of isospectrality exists in that case). The equations of motion following from the action~\eqref{eq:EFTAction} are
\begin{equation}\label{eq:HD_Eqs}
    G_{ab}=\ell_\text{\tiny UV}^{4} P^{(6)}_{ab}+\ell_\text{\tiny UV}^{6}P^{(8)}_{ab}\, ,
\end{equation}
where $G_{ab}$ denotes the Einstein tensor, while  $P^{(6)}_{ab}$ and $P^{ (8)}_{ab}$ follow from the six- and eight-derivative corrections, respectively. To study dynamical fluctuations about \eqref{eq:PLLR}, we expand the linear perturbation of the metric as
\begin{equation}
    h_{ab}=h^\text{\tiny GR}_{ab}+\ell_\text{\tiny UV}^{4}\ h^\text{\tiny HD,6}_{ab}+\ell_\text{\tiny UV}^{6} h^\text{\tiny HD,8}_{ab}\, ,
\end{equation}
where the superscript ``GR'' denotes the GR solution to the linearized equations, while ``HD'' indicates the higher-derivative correction due to each of the higher-derivative operators (likewise, linear perturbations of any other quantity are written as $\dot{\Psi}=\dot{\Psi}^\text{\tiny GR}+\ell_\text{\tiny UV}^{4} \ \dot{\Psi}^\text{\tiny HD,6}+\ell_\text{\tiny UV}^{6} \ \dot{\Psi}^\text{\tiny HD,8}$). The equations are obtained by linearizing \eqref{eq:HD_Eqs}, and imposing they hold order by order in $\ell_\text{\tiny UV}$. At zeroth, fourth and sixth order in $\ell_\text{\tiny UV}$ they give $\dot{G}_{ab}[h^\text{\tiny GR}]=0$, $\dot{G}_{ab}[h^\text{\tiny HD,6}]=\dot{P}^{(6)}_{ab}[h^\text{\tiny GR}]$ and $\dot{G}_{ab}[h^\text{\tiny HD,8}]=\dot{P}^{(8)}_{ab}[h^\text{\tiny GR}]$. However, a tedious but straightforward computation reveals that, if $h^\text{\tiny GR}$ in Eq.~\eqref{eq:recmet2_final} is on-shell, then $\dot P^{(6)}_{ab}\left[h^\text{\tiny GR}\right]=0$ and the six-derivative term induces no correction to the dynamics, $h^\text{\tiny HD,6}=0$ (see also Ref.~\cite{Gruzinov:2006ie}). The eight-derivative term, however, does. For the extremal-weight Weyl scalar $\dot{\Psi}_{0}$, we obtain
\begin{equation}
\begin{aligned}\label{eq:Psi0_HD}
    \square \dot{\Psi}^\text{\tiny HD,8}_{0}&=-64 p_{v}^{4}\Omega^{4}\left\{\left(\tilde{\epsilon}_{1}+\tilde{\epsilon}_{2}\right)\dot{\Psi}^\text{\tiny GR}_{0}+\left(\tilde{\epsilon}_{1}-\tilde{\epsilon}_{2}\right)\dot{\bar{\Psi}}^\text{\tiny GR}_{0}\right\}\, ,
\end{aligned}
\end{equation}
where $\tilde{\epsilon}_{1,2}=\epsilon_{1,2}/\ell_\text{\tiny UV}^{6}$ are the dimensionless couplings. Similar equations can be obtained for the remaining Weyl scalars as well as other GHP quantities. Requiring that the corrected (anti-)self-dual Weyl tensor transforms as $\dot{C}^{\pm,\text{\tiny HD}}\mapsto e^{\pm i\theta}\dot{C}^{\pm,\text{\tiny HD}}$ under \eqref{eq:EMDualityHertz_final} forbids the occurrence of $\dot{\bar{\Psi}}^\text{\tiny GR}_{0}$ in the source terms of~\eqref{eq:Psi0_HD}, and allows only for $\dot{\Psi}^\text{\tiny GR}_{0}$, which transforms with the correct sign under~\eqref{eq:EMDualityHertz_final}. That is, we find
\begin{equation}\label{eq:Iso_couplings}
 \dot{C}^{\pm,\text{\tiny HD}}\mapsto e^{\pm i\theta}\dot{C}^{\pm,\text{\tiny HD}} \qquad \implies \qquad    \epsilon_{1}=\epsilon_{2}\equiv \epsilon\, .
\end{equation}
The explicit expressions supporting this argument are given in the SM. In parallel, Eq.~\eqref{eq:Iso_couplings} turns out to be precisely the combination that preserves isospectrality: Eq.~\eqref{eq:Psi0_HD} for the pairs $\text{Re}\left[\dot{\Psi}^\text{\tiny HD,8}_{0}\right],\text{Re}\left[\dot{\Psi}^\text{\tiny GR}_{0}\right]$ and $\text{Im}\left[\dot{\Psi}^\text{\tiny HD,8}_{0}\right],\text{Im}\left[\dot{\Psi}^\text{\tiny GR}_{0}\right]$ are identical only if \eqref{eq:Iso_couplings} holds.

The shift in the frequencies can be obtained straightforwardly,
\begin{equation}\label{eq:QNM_Shift}
    \delta p_{u}=64\Omega^{4}p_{v}^{3}\epsilon\,,
\end{equation}
providing a universal correction to both sectors (see SM for details).

\vspace{0.1cm}
\noindent{{\bf{Discussion.}}} 
In this work we traced the geometric origin of eikonal isospectrality to the local light ring geometry. We showed that gravitational electric-magnetic duality is an on-shell symmetry of generic perturbations on Penrose-limit pp-waves, both in GR and in a certain class of eight-derivative curvature corrections, and that this symmetry is responsible for the isospectrality of the pp-wave characteristic modes in those theories. Because duality transformations mix polarizations, duality invariance also guarantees nonbirefringent propagation (see~\cite{Okounkova:2021xjv} for observational probes).

In GR, the characteristic modes of these pp-waves correspond to the eikonal regime of black hole QNMs~\cite{Fransen:2023eqj,Fransen:2025cgv,Perrone:2025zhy,Kehagias:2025ntm}, so our results apply directly to that setting. This indicates that eikonal QNM isospectrality and the absence of birefringence in GR in that regime stem from the duality invariance we have identified. When higher-derivative corrections are present, the correspondence between Penrose limits and eikonal QNMs has not yet been established; our findings suggest that such a correspondence should hold. 

Working on Penrose-limit pp-waves also allows us to recover key results previously obtained for the eikonal regime of EFT-corrected black hole QNMs (see Ref.~\cite{Cano:2024wzo}). In particular, we recover isospectrality only if $\epsilon_{1}=\epsilon_{2}\equiv\epsilon$ --- precisely the curvature corrections predicted by type II string theory~\cite{Schwarz:1995dk}. In addition, in agreement with Refs.~\cite{Cano:2024wzo,Cano:2025mht}, we find that the correction to the real part of the QNM frequencies scales as $\delta\omega_\text{\tiny R}\sim \epsilon \, \ell^{3}$ [see Eq.~\eqref{eq:QNM_Shift}], upon identifying the angular multipole moment $p_{v}\sim\ell$, as dictated by the pp-wave and eikonal correspondence~\cite{Fransen:2023eqj}. We also obtain the absence of birefringence,  which was also found in~\cite{Cano:2024wzo}. On the other hand, our framework is not able to recover neither the constraint $\lambda_{\text{e}}=0$ (since the dynamics of a pp-wave are unaffected by cubic terms) nor the corrections to the imaginary part of the QNMs. In fact, in~\cite{Cano:2024wzo,Cano:2025mht} it was found that both the cubic corrections to QNMs, as well as corrections in the damping times, are next-to-leading order effects in the eikonal expansion, while shifts to the (real) frequency appear at leading order. This leads us to conjecture that the correspondence between pp-wave perturbations and the eikonal regime of QNMs extends to EFT modifications of GR: our framework  should capture the leading-order effects, while refinements will be required to account for subleading corrections.

Our approach thus ties the symmetry of the theory (gravitational duality in the eikonal limit) directly to observables, including eikonal QNMs. As a result, one can identify isospectrality-preserving corrections already at the Lagrangian level by allowing only terms that respect that duality --- notice that invariance of \eqref{eq:EFTAction} under $C^{\pm}\mapsto e^{\pm i \theta}C^{\pm}$ yields not only $\epsilon_{1}=\epsilon_{2}$, but also $\lambda_{\text{e}}=0$. This criterion is a gravitational analog to duality constraints in nonlinear interacting electrodynamics, including the Born-Infeld and Gaillard-Zumino theories~\cite{Gaillard:1981rj,Gibbons:1995cv,Novotny:2018iph, Rosly:2002jt}, as well as some supergravity matter models~\cite{Pope:2025jgz}. This opens the door to constraining  symmetries of gravitational theories from observables such as QNMs. While the effects are likely very small in most situations, there are regimes where they could become relevant, as argued in~\cite{Cano:2024wzo}.  

Our work can be extended in several directions. A primary task is to clarify the correspondence between the Penrose limit and the eikonal regime of black holes beyond GR, which would let us include next-to-leading-order eikonal corrections. Extremal black holes are a particularly interesting laboratory for this, since they are highly sensitive to curvature corrections~\cite{Cano:2025mht, Boyce:2026rnn}. We also expect the viewpoint developed here to provide a concrete link between QNM isospectrality and the gravitational duality encountered in other high-energy contexts~\cite{Hull:2000rr,Hull:2001iu}. We plan to explore these directions in future work.

\vspace{0.1cm}
\noindent{{\bf Acknowledgments.}}
We thank Yanbei Chen, Nicola Franchini and Hector O.~Silva for interesting discussions. 
I.B. and B.G. are supported in part by the Simons Collaboration
on Global Categorical Symmetries and also by the NSF Grant No.~PHY-2412361.
E.B., V.D.L., and D.P. are supported by NSF Grants No.~AST-2307146, No.~PHY-2513337, No.~PHY-090003, and No.~PHY-20043, by NASA Grant No.~21-ATP21-0010, by John Templeton Foundation Grant No.~62840, by the Simons Foundation [MPS-SIP-00001698, E.B.], by the Simons Foundation International [SFI-MPS-BH-00012593-02], and by Italian Ministry of Foreign Affairs and International Cooperation Grant No.~PGR01167.
This work was carried out at the Advanced Research Computing at Hopkins (ARCH) core facility (\url{https://www.arch.jhu.edu/}), which is supported by the NSF Grant No.~OAC-1920103.

\bibliography{ref}

\begin{thebibliography}{89}%
\makeatletter
\providecommand \@ifxundefined [1]{%
 \@ifx{#1\undefined}
}%
\providecommand \@ifnum [1]{%
 \ifnum #1\expandafter \@firstoftwo
 \else \expandafter \@secondoftwo
 \fi
}%
\providecommand \@ifx [1]{%
 \ifx #1\expandafter \@firstoftwo
 \else \expandafter \@secondoftwo
 \fi
}%
\providecommand \natexlab [1]{#1}%
\providecommand \enquote  [1]{``#1''}%
\providecommand \bibnamefont  [1]{#1}%
\providecommand \bibfnamefont [1]{#1}%
\providecommand \citenamefont [1]{#1}%
\providecommand \href@noop [0]{\@secondoftwo}%
\providecommand \href [0]{\begingroup \@sanitize@url \@href}%
\providecommand \@href[1]{\@@startlink{#1}\@@href}%
\providecommand \@@href[1]{\endgroup#1\@@endlink}%
\providecommand \@sanitize@url [0]{\catcode `\\12\catcode `\$12\catcode
  `\&12\catcode `\#12\catcode `\^12\catcode `\_12\catcode `\%12\relax}%
\providecommand \@@startlink[1]{}%
\providecommand \@@endlink[0]{}%
\providecommand \url  [0]{\begingroup\@sanitize@url \@url }%
\providecommand \@url [1]{\endgroup\@href {#1}{\urlprefix }}%
\providecommand \urlprefix  [0]{URL }%
\providecommand \Eprint [0]{\href }%
\providecommand \doibase [0]{https://doi.org/}%
\providecommand \selectlanguage [0]{\@gobble}%
\providecommand \bibinfo  [0]{\@secondoftwo}%
\providecommand \bibfield  [0]{\@secondoftwo}%
\providecommand \translation [1]{[#1]}%
\providecommand \BibitemOpen [0]{}%
\providecommand \bibitemStop [0]{}%
\providecommand \bibitemNoStop [0]{.\EOS\space}%
\providecommand \EOS [0]{\spacefactor3000\relax}%
\providecommand \BibitemShut  [1]{\csname bibitem#1\endcsname}%
\let\auto@bib@innerbib\@empty
\bibitem [{\citenamefont {Berti}\ \emph {et~al.}(2025)\citenamefont {Berti},
  \citenamefont {Cardoso}, \citenamefont {Carullo} \emph
  {et~al.}}]{Berti:2025hly}%
  \BibitemOpen
  \bibfield  {author} {\bibinfo {author} {\bibfnamefont {E.}~\bibnamefont
  {Berti}}, \bibinfo {author} {\bibfnamefont {V.}~\bibnamefont {Cardoso}},
  \bibinfo {author} {\bibfnamefont {G.}~\bibnamefont {Carullo}}, \emph
  {et~al.},\ }\bibfield  {title} {\bibinfo {title} {{Black hole spectroscopy:
  from theory to experiment}},\ }\href@noop {} {\  (\bibinfo {year} {2025})},\
  \Eprint {https://arxiv.org/abs/2505.23895} {arXiv:2505.23895 [gr-qc]}
  \BibitemShut {NoStop}%
\bibitem [{\citenamefont {Vishveshwara}(1970)}]{Vishveshwara:1970zz}%
  \BibitemOpen
  \bibfield  {author} {\bibinfo {author} {\bibfnamefont {C.~V.}\ \bibnamefont
  {Vishveshwara}},\ }\bibfield  {title} {\bibinfo {title} {{Scattering of
  Gravitational Radiation by a Schwarzschild Black-hole}},\ }\href
  {https://doi.org/10.1038/227936a0} {\bibfield  {journal} {\bibinfo  {journal}
  {Nature}\ }\textbf {\bibinfo {volume} {227}},\ \bibinfo {pages} {936}
  (\bibinfo {year} {1970})}\BibitemShut {NoStop}%
\bibitem [{\citenamefont {Press}(1971)}]{Press:1971wr}%
  \BibitemOpen
  \bibfield  {author} {\bibinfo {author} {\bibfnamefont {W.~H.}\ \bibnamefont
  {Press}},\ }\bibfield  {title} {\bibinfo {title} {{Long Wave Trains of
  Gravitational Waves from a Vibrating Black Hole}},\ }\href
  {https://doi.org/10.1086/180849} {\bibfield  {journal} {\bibinfo  {journal}
  {Astrophys. J. Lett.}\ }\textbf {\bibinfo {volume} {170}},\ \bibinfo {pages}
  {L105} (\bibinfo {year} {1971})}\BibitemShut {NoStop}%
\bibitem [{\citenamefont {Teukolsky}(1973)}]{Teukolsky:1973ha}%
  \BibitemOpen
  \bibfield  {author} {\bibinfo {author} {\bibfnamefont {S.~A.}\ \bibnamefont
  {Teukolsky}},\ }\bibfield  {title} {\bibinfo {title} {{Perturbations of a
  rotating black hole. 1. Fundamental equations for gravitational
  electromagnetic and neutrino field perturbations}},\ }\href
  {https://doi.org/10.1086/152444} {\bibfield  {journal} {\bibinfo  {journal}
  {Astrophys. J.}\ }\textbf {\bibinfo {volume} {185}},\ \bibinfo {pages} {635}
  (\bibinfo {year} {1973})}\BibitemShut {NoStop}%
\bibitem [{\citenamefont {Chandrasekhar}\ and\ \citenamefont
  {Detweiler}(1975)}]{Chandrasekhar:1975zza}%
  \BibitemOpen
  \bibfield  {author} {\bibinfo {author} {\bibfnamefont {S.}~\bibnamefont
  {Chandrasekhar}}\ and\ \bibinfo {author} {\bibfnamefont {S.~L.}\ \bibnamefont
  {Detweiler}},\ }\bibfield  {title} {\bibinfo {title} {{The quasi-normal modes
  of the Schwarzschild black hole}},\ }\href
  {https://doi.org/10.1098/rspa.1975.0112} {\bibfield  {journal} {\bibinfo
  {journal} {Proc. Roy. Soc. Lond. A}\ }\textbf {\bibinfo {volume} {344}},\
  \bibinfo {pages} {441} (\bibinfo {year} {1975})}\BibitemShut {NoStop}%
\bibitem [{\citenamefont {Kokkotas}\ and\ \citenamefont
  {Schmidt}(1999)}]{Kokkotas:1999bd}%
  \BibitemOpen
  \bibfield  {author} {\bibinfo {author} {\bibfnamefont {K.~D.}\ \bibnamefont
  {Kokkotas}}\ and\ \bibinfo {author} {\bibfnamefont {B.~G.}\ \bibnamefont
  {Schmidt}},\ }\bibfield  {title} {\bibinfo {title} {{Quasinormal modes of
  stars and black holes}},\ }\href {https://doi.org/10.12942/lrr-1999-2}
  {\bibfield  {journal} {\bibinfo  {journal} {Living Rev. Rel.}\ }\textbf
  {\bibinfo {volume} {2}},\ \bibinfo {pages} {2} (\bibinfo {year} {1999})},\
  \Eprint {https://arxiv.org/abs/gr-qc/9909058} {arXiv:gr-qc/9909058}
  \BibitemShut {NoStop}%
\bibitem [{\citenamefont {Berti}\ \emph {et~al.}(2009)\citenamefont {Berti},
  \citenamefont {Cardoso},\ and\ \citenamefont {Starinets}}]{Berti:2009kk}%
  \BibitemOpen
  \bibfield  {author} {\bibinfo {author} {\bibfnamefont {E.}~\bibnamefont
  {Berti}}, \bibinfo {author} {\bibfnamefont {V.}~\bibnamefont {Cardoso}},\
  and\ \bibinfo {author} {\bibfnamefont {A.~O.}\ \bibnamefont {Starinets}},\
  }\bibfield  {title} {\bibinfo {title} {{Quasinormal modes of black holes and
  black branes}},\ }\href {https://doi.org/10.1088/0264-9381/26/16/163001}
  {\bibfield  {journal} {\bibinfo  {journal} {Class. Quant. Grav.}\ }\textbf
  {\bibinfo {volume} {26}},\ \bibinfo {pages} {163001} (\bibinfo {year}
  {2009})},\ \Eprint {https://arxiv.org/abs/0905.2975} {arXiv:0905.2975
  [gr-qc]} \BibitemShut {NoStop}%
\bibitem [{\citenamefont {Dreyer}\ \emph {et~al.}(2004)\citenamefont {Dreyer},
  \citenamefont {Kelly}, \citenamefont {Krishnan}, \citenamefont {Finn},
  \citenamefont {Garrison},\ and\ \citenamefont
  {Lopez-Aleman}}]{Dreyer:2003bv}%
  \BibitemOpen
  \bibfield  {author} {\bibinfo {author} {\bibfnamefont {O.}~\bibnamefont
  {Dreyer}}, \bibinfo {author} {\bibfnamefont {B.~J.}\ \bibnamefont {Kelly}},
  \bibinfo {author} {\bibfnamefont {B.}~\bibnamefont {Krishnan}}, \bibinfo
  {author} {\bibfnamefont {L.~S.}\ \bibnamefont {Finn}}, \bibinfo {author}
  {\bibfnamefont {D.}~\bibnamefont {Garrison}},\ and\ \bibinfo {author}
  {\bibfnamefont {R.}~\bibnamefont {Lopez-Aleman}},\ }\bibfield  {title}
  {\bibinfo {title} {{Black hole spectroscopy: Testing general relativity
  through gravitational wave observations}},\ }\href
  {https://doi.org/10.1088/0264-9381/21/4/003} {\bibfield  {journal} {\bibinfo
  {journal} {Class. Quant. Grav.}\ }\textbf {\bibinfo {volume} {21}},\ \bibinfo
  {pages} {787} (\bibinfo {year} {2004})},\ \Eprint
  {https://arxiv.org/abs/gr-qc/0309007} {arXiv:gr-qc/0309007} \BibitemShut
  {NoStop}%
\bibitem [{\citenamefont {Berti}\ \emph {et~al.}(2006)\citenamefont {Berti},
  \citenamefont {Cardoso},\ and\ \citenamefont {Will}}]{Berti:2005ys}%
  \BibitemOpen
  \bibfield  {author} {\bibinfo {author} {\bibfnamefont {E.}~\bibnamefont
  {Berti}}, \bibinfo {author} {\bibfnamefont {V.}~\bibnamefont {Cardoso}},\
  and\ \bibinfo {author} {\bibfnamefont {C.~M.}\ \bibnamefont {Will}},\
  }\bibfield  {title} {\bibinfo {title} {{On gravitational-wave spectroscopy of
  massive black holes with the space interferometer LISA}},\ }\href
  {https://doi.org/10.1103/PhysRevD.73.064030} {\bibfield  {journal} {\bibinfo
  {journal} {Phys. Rev. D}\ }\textbf {\bibinfo {volume} {73}},\ \bibinfo
  {pages} {064030} (\bibinfo {year} {2006})},\ \Eprint
  {https://arxiv.org/abs/gr-qc/0512160} {arXiv:gr-qc/0512160} \BibitemShut
  {NoStop}%
\bibitem [{\citenamefont {Gossan}\ \emph {et~al.}(2012)\citenamefont {Gossan},
  \citenamefont {Veitch},\ and\ \citenamefont {Sathyaprakash}}]{Gossan:2011ha}%
  \BibitemOpen
  \bibfield  {author} {\bibinfo {author} {\bibfnamefont {S.}~\bibnamefont
  {Gossan}}, \bibinfo {author} {\bibfnamefont {J.}~\bibnamefont {Veitch}},\
  and\ \bibinfo {author} {\bibfnamefont {B.~S.}\ \bibnamefont
  {Sathyaprakash}},\ }\bibfield  {title} {\bibinfo {title} {{Bayesian model
  selection for testing the no-hair theorem with black hole ringdowns}},\
  }\href {https://doi.org/10.1103/PhysRevD.85.124056} {\bibfield  {journal}
  {\bibinfo  {journal} {Phys. Rev. D}\ }\textbf {\bibinfo {volume} {85}},\
  \bibinfo {pages} {124056} (\bibinfo {year} {2012})},\ \Eprint
  {https://arxiv.org/abs/1111.5819} {arXiv:1111.5819 [gr-qc]} \BibitemShut
  {NoStop}%
\bibitem [{\citenamefont {Berti}\ \emph {et~al.}(2015)\citenamefont {Berti}
  \emph {et~al.}}]{Berti:2015itd}%
  \BibitemOpen
  \bibfield  {author} {\bibinfo {author} {\bibfnamefont {E.}~\bibnamefont
  {Berti}} \emph {et~al.},\ }\bibfield  {title} {\bibinfo {title} {{Testing
  General Relativity with Present and Future Astrophysical Observations}},\
  }\href {https://doi.org/10.1088/0264-9381/32/24/243001} {\bibfield  {journal}
  {\bibinfo  {journal} {Class. Quant. Grav.}\ }\textbf {\bibinfo {volume}
  {32}},\ \bibinfo {pages} {243001} (\bibinfo {year} {2015})},\ \Eprint
  {https://arxiv.org/abs/1501.07274} {arXiv:1501.07274 [gr-qc]} \BibitemShut
  {NoStop}%
\bibitem [{\citenamefont {Berti}\ \emph {et~al.}(2018)\citenamefont {Berti},
  \citenamefont {Yagi}, \citenamefont {Yang},\ and\ \citenamefont
  {Yunes}}]{Berti:2018vdi}%
  \BibitemOpen
  \bibfield  {author} {\bibinfo {author} {\bibfnamefont {E.}~\bibnamefont
  {Berti}}, \bibinfo {author} {\bibfnamefont {K.}~\bibnamefont {Yagi}},
  \bibinfo {author} {\bibfnamefont {H.}~\bibnamefont {Yang}},\ and\ \bibinfo
  {author} {\bibfnamefont {N.}~\bibnamefont {Yunes}},\ }\bibfield  {title}
  {\bibinfo {title} {{Extreme Gravity Tests with Gravitational Waves from
  Compact Binary Coalescences: (II) Ringdown}},\ }\href
  {https://doi.org/10.1007/s10714-018-2372-6} {\bibfield  {journal} {\bibinfo
  {journal} {Gen. Rel. Grav.}\ }\textbf {\bibinfo {volume} {50}},\ \bibinfo
  {pages} {49} (\bibinfo {year} {2018})},\ \Eprint
  {https://arxiv.org/abs/1801.03587} {arXiv:1801.03587 [gr-qc]} \BibitemShut
  {NoStop}%
\bibitem [{\citenamefont {Franchini}\ and\ \citenamefont
  {V{\"o}lkel}(2024)}]{Franchini:2023eda}%
  \BibitemOpen
  \bibfield  {author} {\bibinfo {author} {\bibfnamefont {N.}~\bibnamefont
  {Franchini}}\ and\ \bibinfo {author} {\bibfnamefont {S.~H.}\ \bibnamefont
  {V{\"o}lkel}},\ }\bibinfo {title} {{Testing General Relativity with Black
  Hole Quasi-normal Modes}}\ (\bibinfo {year} {2024})\ \Eprint
  {https://arxiv.org/abs/2305.01696} {arXiv:2305.01696 [gr-qc]} \BibitemShut
  {NoStop}%
\bibitem [{\citenamefont {Abac}\ \emph
  {et~al.}(2026{\natexlab{a}})\citenamefont {Abac} \emph
  {et~al.}}]{LIGOScientific:2026qni}%
  \BibitemOpen
  \bibfield  {author} {\bibinfo {author} {\bibfnamefont {A.~G.}\ \bibnamefont
  {Abac}} \emph {et~al.} (\bibinfo {collaboration} {LIGO Scientific, VIRGO,
  KAGRA}),\ }\bibfield  {title} {\bibinfo {title} {{GWTC-4.0: Tests of General
  Relativity. I. Overview and General Tests}},\ }\href@noop {} {\  (\bibinfo
  {year} {2026}{\natexlab{a}})},\ \Eprint {https://arxiv.org/abs/2603.19019}
  {arXiv:2603.19019 [gr-qc]} \BibitemShut {NoStop}%
\bibitem [{\citenamefont {Abac}\ \emph
  {et~al.}(2026{\natexlab{b}})\citenamefont {Abac} \emph
  {et~al.}}]{LIGOScientific:2026wpt}%
  \BibitemOpen
  \bibfield  {author} {\bibinfo {author} {\bibfnamefont {A.~G.}\ \bibnamefont
  {Abac}} \emph {et~al.} (\bibinfo {collaboration} {LIGO Scientific, VIRGO,
  KAGRA}),\ }\bibfield  {title} {\bibinfo {title} {{GWTC-4.0: Tests of General
  Relativity. III. Tests of the Remnants}},\ }\href@noop {} {\  (\bibinfo
  {year} {2026}{\natexlab{b}})},\ \Eprint {https://arxiv.org/abs/2603.19021}
  {arXiv:2603.19021 [gr-qc]} \BibitemShut {NoStop}%
\bibitem [{\citenamefont {Regge}\ and\ \citenamefont
  {Wheeler}(1957)}]{Regge:1957td}%
  \BibitemOpen
  \bibfield  {author} {\bibinfo {author} {\bibfnamefont {T.}~\bibnamefont
  {Regge}}\ and\ \bibinfo {author} {\bibfnamefont {J.~A.}\ \bibnamefont
  {Wheeler}},\ }\bibfield  {title} {\bibinfo {title} {{Stability of a
  Schwarzschild singularity}},\ }\href
  {https://doi.org/10.1103/PhysRev.108.1063} {\bibfield  {journal} {\bibinfo
  {journal} {Phys. Rev.}\ }\textbf {\bibinfo {volume} {108}},\ \bibinfo {pages}
  {1063} (\bibinfo {year} {1957})}\BibitemShut {NoStop}%
\bibitem [{\citenamefont {Zerilli}(1970)}]{Zerilli:1970se}%
  \BibitemOpen
  \bibfield  {author} {\bibinfo {author} {\bibfnamefont {F.~J.}\ \bibnamefont
  {Zerilli}},\ }\bibfield  {title} {\bibinfo {title} {{Effective potential for
  even parity Regge-Wheeler gravitational perturbation equations}},\ }\href
  {https://doi.org/10.1103/PhysRevLett.24.737} {\bibfield  {journal} {\bibinfo
  {journal} {Phys. Rev. Lett.}\ }\textbf {\bibinfo {volume} {24}},\ \bibinfo
  {pages} {737} (\bibinfo {year} {1970})}\BibitemShut {NoStop}%
\bibitem [{\citenamefont {Chandrasekhar}(1975)}]{Chandrasekhar:1975nkd}%
  \BibitemOpen
  \bibfield  {author} {\bibinfo {author} {\bibfnamefont {S.}~\bibnamefont
  {Chandrasekhar}},\ }\bibfield  {title} {\bibinfo {title} {{On the equations
  governing the perturbations of the Schwarzschild black hole}},\ }\href
  {https://doi.org/10.1098/rspa.1975.0066} {\bibfield  {journal} {\bibinfo
  {journal} {Proc. Roy. Soc. Lond. A}\ }\textbf {\bibinfo {volume} {343}},\
  \bibinfo {pages} {289} (\bibinfo {year} {1975})}\BibitemShut {NoStop}%
\bibitem [{\citenamefont {Kallosh}\ \emph {et~al.}(1998)\citenamefont
  {Kallosh}, \citenamefont {Rahmfeld},\ and\ \citenamefont
  {Wong}}]{Kallosh:1997ug}%
  \BibitemOpen
  \bibfield  {author} {\bibinfo {author} {\bibfnamefont {R.}~\bibnamefont
  {Kallosh}}, \bibinfo {author} {\bibfnamefont {J.}~\bibnamefont {Rahmfeld}},\
  and\ \bibinfo {author} {\bibfnamefont {W.~K.}\ \bibnamefont {Wong}},\
  }\bibfield  {title} {\bibinfo {title} {{One loop supergravity corrections to
  the black hole entropy and residual supersymmetry}},\ }\href
  {https://doi.org/10.1103/PhysRevD.57.1063} {\bibfield  {journal} {\bibinfo
  {journal} {Phys. Rev. D}\ }\textbf {\bibinfo {volume} {57}},\ \bibinfo
  {pages} {1063} (\bibinfo {year} {1998})},\ \Eprint
  {https://arxiv.org/abs/hep-th/9706048} {arXiv:hep-th/9706048} \BibitemShut
  {NoStop}%
\bibitem [{\citenamefont {Glampedakis}\ \emph {et~al.}(2017)\citenamefont
  {Glampedakis}, \citenamefont {Johnson},\ and\ \citenamefont
  {Kennefick}}]{Glampedakis:2017rar}%
  \BibitemOpen
  \bibfield  {author} {\bibinfo {author} {\bibfnamefont {K.}~\bibnamefont
  {Glampedakis}}, \bibinfo {author} {\bibfnamefont {A.~D.}\ \bibnamefont
  {Johnson}},\ and\ \bibinfo {author} {\bibfnamefont {D.}~\bibnamefont
  {Kennefick}},\ }\bibfield  {title} {\bibinfo {title} {{Darboux transformation
  in black hole perturbation theory}},\ }\href
  {https://doi.org/10.1103/PhysRevD.96.024036} {\bibfield  {journal} {\bibinfo
  {journal} {Phys. Rev. D}\ }\textbf {\bibinfo {volume} {96}},\ \bibinfo
  {pages} {024036} (\bibinfo {year} {2017})},\ \Eprint
  {https://arxiv.org/abs/1702.06459} {arXiv:1702.06459 [gr-qc]} \BibitemShut
  {NoStop}%
\bibitem [{\citenamefont {Lenzi}\ and\ \citenamefont
  {Sopuerta}(2021{\natexlab{a}})}]{Lenzi:2021njy}%
  \BibitemOpen
  \bibfield  {author} {\bibinfo {author} {\bibfnamefont {M.}~\bibnamefont
  {Lenzi}}\ and\ \bibinfo {author} {\bibfnamefont {C.~F.}\ \bibnamefont
  {Sopuerta}},\ }\bibfield  {title} {\bibinfo {title} {{Darboux covariance: A
  hidden symmetry of perturbed Schwarzschild black holes}},\ }\href
  {https://doi.org/10.1103/PhysRevD.104.124068} {\bibfield  {journal} {\bibinfo
   {journal} {Phys. Rev. D}\ }\textbf {\bibinfo {volume} {104}},\ \bibinfo
  {pages} {124068} (\bibinfo {year} {2021}{\natexlab{a}})},\ \Eprint
  {https://arxiv.org/abs/2109.00503} {arXiv:2109.00503 [gr-qc]} \BibitemShut
  {NoStop}%
\bibitem [{\citenamefont {Lenzi}\ and\ \citenamefont
  {Sopuerta}(2021{\natexlab{b}})}]{Lenzi:2021wpc}%
  \BibitemOpen
  \bibfield  {author} {\bibinfo {author} {\bibfnamefont {M.}~\bibnamefont
  {Lenzi}}\ and\ \bibinfo {author} {\bibfnamefont {C.~F.}\ \bibnamefont
  {Sopuerta}},\ }\bibfield  {title} {\bibinfo {title} {{Master functions and
  equations for perturbations of vacuum spherically symmetric spacetimes}},\
  }\href {https://doi.org/10.1103/PhysRevD.104.084053} {\bibfield  {journal}
  {\bibinfo  {journal} {Phys. Rev. D}\ }\textbf {\bibinfo {volume} {104}},\
  \bibinfo {pages} {084053} (\bibinfo {year} {2021}{\natexlab{b}})},\ \Eprint
  {https://arxiv.org/abs/2108.08668} {arXiv:2108.08668 [gr-qc]} \BibitemShut
  {NoStop}%
\bibitem [{\citenamefont {Lenzi}\ and\ \citenamefont
  {Sopuerta}(2023{\natexlab{a}})}]{Lenzi:2022wjv}%
  \BibitemOpen
  \bibfield  {author} {\bibinfo {author} {\bibfnamefont {M.}~\bibnamefont
  {Lenzi}}\ and\ \bibinfo {author} {\bibfnamefont {C.~F.}\ \bibnamefont
  {Sopuerta}},\ }\bibfield  {title} {\bibinfo {title} {{Black hole greybody
  factors from Korteweg{\textendash}de Vries integrals: Theory}},\ }\href
  {https://doi.org/10.1103/PhysRevD.107.044010} {\bibfield  {journal} {\bibinfo
   {journal} {Phys. Rev. D}\ }\textbf {\bibinfo {volume} {107}},\ \bibinfo
  {pages} {044010} (\bibinfo {year} {2023}{\natexlab{a}})},\ \Eprint
  {https://arxiv.org/abs/2212.03732} {arXiv:2212.03732 [gr-qc]} \BibitemShut
  {NoStop}%
\bibitem [{\citenamefont {Lenzi}\ and\ \citenamefont
  {Sopuerta}(2023{\natexlab{b}})}]{Lenzi:2023inn}%
  \BibitemOpen
  \bibfield  {author} {\bibinfo {author} {\bibfnamefont {M.}~\bibnamefont
  {Lenzi}}\ and\ \bibinfo {author} {\bibfnamefont {C.~F.}\ \bibnamefont
  {Sopuerta}},\ }\bibfield  {title} {\bibinfo {title} {{Black hole greybody
  factors from Korteweg{\textendash}de Vries integrals: Computation}},\ }\href
  {https://doi.org/10.1103/PhysRevD.107.084039} {\bibfield  {journal} {\bibinfo
   {journal} {Phys. Rev. D}\ }\textbf {\bibinfo {volume} {107}},\ \bibinfo
  {pages} {084039} (\bibinfo {year} {2023}{\natexlab{b}})},\ \Eprint
  {https://arxiv.org/abs/2301.01096} {arXiv:2301.01096 [gr-qc]} \BibitemShut
  {NoStop}%
\bibitem [{\citenamefont {Solomon}(2023)}]{Solomon:2023ltn}%
  \BibitemOpen
  \bibfield  {author} {\bibinfo {author} {\bibfnamefont {A.~R.}\ \bibnamefont
  {Solomon}},\ }\bibfield  {title} {\bibinfo {title} {{Off-Shell Duality
  Invariance of Schwarzschild Perturbation Theory}},\ }\href
  {https://doi.org/10.3390/particles6040061} {\bibfield  {journal} {\bibinfo
  {journal} {Particles}\ }\textbf {\bibinfo {volume} {6}},\ \bibinfo {pages}
  {943} (\bibinfo {year} {2023})},\ \Eprint {https://arxiv.org/abs/2310.04502}
  {arXiv:2310.04502 [gr-qc]} \BibitemShut {NoStop}%
\bibitem [{\citenamefont {Nichols}\ \emph {et~al.}(2012)\citenamefont
  {Nichols}, \citenamefont {Zimmerman}, \citenamefont {Chen}, \citenamefont
  {Lovelace}, \citenamefont {Matthews}, \citenamefont {Owen}, \citenamefont
  {Zhang},\ and\ \citenamefont {Thorne}}]{Nichols:2012jn}%
  \BibitemOpen
  \bibfield  {author} {\bibinfo {author} {\bibfnamefont {D.~A.}\ \bibnamefont
  {Nichols}}, \bibinfo {author} {\bibfnamefont {A.}~\bibnamefont {Zimmerman}},
  \bibinfo {author} {\bibfnamefont {Y.}~\bibnamefont {Chen}}, \bibinfo {author}
  {\bibfnamefont {G.}~\bibnamefont {Lovelace}}, \bibinfo {author}
  {\bibfnamefont {K.~D.}\ \bibnamefont {Matthews}}, \bibinfo {author}
  {\bibfnamefont {R.}~\bibnamefont {Owen}}, \bibinfo {author} {\bibfnamefont
  {F.}~\bibnamefont {Zhang}},\ and\ \bibinfo {author} {\bibfnamefont {K.~S.}\
  \bibnamefont {Thorne}},\ }\bibfield  {title} {\bibinfo {title} {{Visualizing
  Spacetime Curvature via Frame-Drag Vortexes and Tidal Tendexes III.
  Quasinormal Pulsations of Schwarzschild and Kerr Black Holes}},\ }\href
  {https://doi.org/10.1103/PhysRevD.86.104028} {\bibfield  {journal} {\bibinfo
  {journal} {Phys. Rev. D}\ }\textbf {\bibinfo {volume} {86}},\ \bibinfo
  {pages} {104028} (\bibinfo {year} {2012})},\ \Eprint
  {https://arxiv.org/abs/1208.3038} {arXiv:1208.3038 [gr-qc]} \BibitemShut
  {NoStop}%
\bibitem [{\citenamefont {Franchini}(2023)}]{Franchini:2023xhd}%
  \BibitemOpen
  \bibfield  {author} {\bibinfo {author} {\bibfnamefont {N.}~\bibnamefont
  {Franchini}},\ }\bibfield  {title} {\bibinfo {title} {{Slow rotation black
  hole perturbation theory}},\ }\href
  {https://doi.org/10.1103/PhysRevD.108.044079} {\bibfield  {journal} {\bibinfo
   {journal} {Phys. Rev. D}\ }\textbf {\bibinfo {volume} {108}},\ \bibinfo
  {pages} {044079} (\bibinfo {year} {2023})},\ \Eprint
  {https://arxiv.org/abs/2305.19313} {arXiv:2305.19313 [gr-qc]} \BibitemShut
  {NoStop}%
\bibitem [{\citenamefont {Mukkamala}\ and\ \citenamefont
  {Pere{\~n}iguez}(2025)}]{Mukkamala:2024dxf}%
  \BibitemOpen
  \bibfield  {author} {\bibinfo {author} {\bibfnamefont {G.~R.}\ \bibnamefont
  {Mukkamala}}\ and\ \bibinfo {author} {\bibfnamefont {D.}~\bibnamefont
  {Pere{\~n}iguez}},\ }\bibfield  {title} {\bibinfo {title} {{Decoupled
  gravitational wave equations in spherical symmetry from curvature wave
  equations}},\ }\href {https://doi.org/10.1088/1475-7516/2025/01/122}
  {\bibfield  {journal} {\bibinfo  {journal} {JCAP}\ }\textbf {\bibinfo
  {volume} {01}},\ \bibinfo {pages} {122}},\ \Eprint
  {https://arxiv.org/abs/2408.13557} {arXiv:2408.13557 [gr-qc]} \BibitemShut
  {NoStop}%
\bibitem [{\citenamefont {Cardoso}\ \emph {et~al.}(2019)\citenamefont
  {Cardoso}, \citenamefont {Kimura}, \citenamefont {Maselli}, \citenamefont
  {Berti}, \citenamefont {Macedo},\ and\ \citenamefont
  {McManus}}]{Cardoso:2019mqo}%
  \BibitemOpen
  \bibfield  {author} {\bibinfo {author} {\bibfnamefont {V.}~\bibnamefont
  {Cardoso}}, \bibinfo {author} {\bibfnamefont {M.}~\bibnamefont {Kimura}},
  \bibinfo {author} {\bibfnamefont {A.}~\bibnamefont {Maselli}}, \bibinfo
  {author} {\bibfnamefont {E.}~\bibnamefont {Berti}}, \bibinfo {author}
  {\bibfnamefont {C.~F.~B.}\ \bibnamefont {Macedo}},\ and\ \bibinfo {author}
  {\bibfnamefont {R.}~\bibnamefont {McManus}},\ }\bibfield  {title} {\bibinfo
  {title} {{Parametrized black hole quasinormal ringdown: Decoupled equations
  for nonrotating black holes}},\ }\href
  {https://doi.org/10.1103/PhysRevD.99.104077} {\bibfield  {journal} {\bibinfo
  {journal} {Phys. Rev. D}\ }\textbf {\bibinfo {volume} {99}},\ \bibinfo
  {pages} {104077} (\bibinfo {year} {2019})},\ \Eprint
  {https://arxiv.org/abs/1901.01265} {arXiv:1901.01265 [gr-qc]} \BibitemShut
  {NoStop}%
\bibitem [{\citenamefont {McManus}\ \emph {et~al.}(2019)\citenamefont
  {McManus}, \citenamefont {Berti}, \citenamefont {Macedo}, \citenamefont
  {Kimura}, \citenamefont {Maselli},\ and\ \citenamefont
  {Cardoso}}]{McManus:2019ulj}%
  \BibitemOpen
  \bibfield  {author} {\bibinfo {author} {\bibfnamefont {R.}~\bibnamefont
  {McManus}}, \bibinfo {author} {\bibfnamefont {E.}~\bibnamefont {Berti}},
  \bibinfo {author} {\bibfnamefont {C.~F.~B.}\ \bibnamefont {Macedo}}, \bibinfo
  {author} {\bibfnamefont {M.}~\bibnamefont {Kimura}}, \bibinfo {author}
  {\bibfnamefont {A.}~\bibnamefont {Maselli}},\ and\ \bibinfo {author}
  {\bibfnamefont {V.}~\bibnamefont {Cardoso}},\ }\bibfield  {title} {\bibinfo
  {title} {{Parametrized black hole quasinormal ringdown. II. Coupled equations
  and quadratic corrections for nonrotating black holes}},\ }\href
  {https://doi.org/10.1103/PhysRevD.100.044061} {\bibfield  {journal} {\bibinfo
   {journal} {Phys. Rev. D}\ }\textbf {\bibinfo {volume} {100}},\ \bibinfo
  {pages} {044061} (\bibinfo {year} {2019})},\ \Eprint
  {https://arxiv.org/abs/1906.05155} {arXiv:1906.05155 [gr-qc]} \BibitemShut
  {NoStop}%
\bibitem [{\citenamefont {Cano}\ \emph {et~al.}(2022)\citenamefont {Cano},
  \citenamefont {Fransen}, \citenamefont {Hertog},\ and\ \citenamefont
  {Maenaut}}]{Cano:2021myl}%
  \BibitemOpen
  \bibfield  {author} {\bibinfo {author} {\bibfnamefont {P.~A.}\ \bibnamefont
  {Cano}}, \bibinfo {author} {\bibfnamefont {K.}~\bibnamefont {Fransen}},
  \bibinfo {author} {\bibfnamefont {T.}~\bibnamefont {Hertog}},\ and\ \bibinfo
  {author} {\bibfnamefont {S.}~\bibnamefont {Maenaut}},\ }\bibfield  {title}
  {\bibinfo {title} {{Gravitational ringing of rotating black holes in
  higher-derivative gravity}},\ }\href
  {https://doi.org/10.1103/PhysRevD.105.024064} {\bibfield  {journal} {\bibinfo
   {journal} {Phys. Rev. D}\ }\textbf {\bibinfo {volume} {105}},\ \bibinfo
  {pages} {024064} (\bibinfo {year} {2022})},\ \Eprint
  {https://arxiv.org/abs/2110.11378} {arXiv:2110.11378 [gr-qc]} \BibitemShut
  {NoStop}%
\bibitem [{\citenamefont {Cano}\ \emph {et~al.}(2023)\citenamefont {Cano},
  \citenamefont {Fransen}, \citenamefont {Hertog},\ and\ \citenamefont
  {Maenaut}}]{Cano:2023jbk}%
  \BibitemOpen
  \bibfield  {author} {\bibinfo {author} {\bibfnamefont {P.~A.}\ \bibnamefont
  {Cano}}, \bibinfo {author} {\bibfnamefont {K.}~\bibnamefont {Fransen}},
  \bibinfo {author} {\bibfnamefont {T.}~\bibnamefont {Hertog}},\ and\ \bibinfo
  {author} {\bibfnamefont {S.}~\bibnamefont {Maenaut}},\ }\bibfield  {title}
  {\bibinfo {title} {{Quasinormal modes of rotating black holes in
  higher-derivative gravity}},\ }\href
  {https://doi.org/10.1103/PhysRevD.108.124032} {\bibfield  {journal} {\bibinfo
   {journal} {Phys. Rev. D}\ }\textbf {\bibinfo {volume} {108}},\ \bibinfo
  {pages} {124032} (\bibinfo {year} {2023})},\ \Eprint
  {https://arxiv.org/abs/2307.07431} {arXiv:2307.07431 [gr-qc]} \BibitemShut
  {NoStop}%
\bibitem [{\citenamefont {Silva}\ \emph {et~al.}(2024)\citenamefont {Silva},
  \citenamefont {Tambalo}, \citenamefont {Glampedakis}, \citenamefont {Yagi},\
  and\ \citenamefont {Steinhoff}}]{Silva:2024ffz}%
  \BibitemOpen
  \bibfield  {author} {\bibinfo {author} {\bibfnamefont {H.~O.}\ \bibnamefont
  {Silva}}, \bibinfo {author} {\bibfnamefont {G.}~\bibnamefont {Tambalo}},
  \bibinfo {author} {\bibfnamefont {K.}~\bibnamefont {Glampedakis}}, \bibinfo
  {author} {\bibfnamefont {K.}~\bibnamefont {Yagi}},\ and\ \bibinfo {author}
  {\bibfnamefont {J.}~\bibnamefont {Steinhoff}},\ }\bibfield  {title} {\bibinfo
  {title} {{Quasinormal modes and their excitation beyond general
  relativity}},\ }\href {https://doi.org/10.1103/PhysRevD.110.024042}
  {\bibfield  {journal} {\bibinfo  {journal} {Phys. Rev. D}\ }\textbf {\bibinfo
  {volume} {110}},\ \bibinfo {pages} {024042} (\bibinfo {year} {2024})},\
  \Eprint {https://arxiv.org/abs/2404.11110} {arXiv:2404.11110 [gr-qc]}
  \BibitemShut {NoStop}%
\bibitem [{\citenamefont {Cano}\ \emph {et~al.}(2024)\citenamefont {Cano},
  \citenamefont {Capuano}, \citenamefont {Franchini}, \citenamefont {Maenaut},\
  and\ \citenamefont {V{\"o}lkel}}]{Cano:2024ezp}%
  \BibitemOpen
  \bibfield  {author} {\bibinfo {author} {\bibfnamefont {P.~A.}\ \bibnamefont
  {Cano}}, \bibinfo {author} {\bibfnamefont {L.}~\bibnamefont {Capuano}},
  \bibinfo {author} {\bibfnamefont {N.}~\bibnamefont {Franchini}}, \bibinfo
  {author} {\bibfnamefont {S.}~\bibnamefont {Maenaut}},\ and\ \bibinfo {author}
  {\bibfnamefont {S.~H.}\ \bibnamefont {V{\"o}lkel}},\ }\bibfield  {title}
  {\bibinfo {title} {{Higher-derivative corrections to the Kerr quasinormal
  mode spectrum}},\ }\href {https://doi.org/10.1103/PhysRevD.110.124057}
  {\bibfield  {journal} {\bibinfo  {journal} {Phys. Rev. D}\ }\textbf {\bibinfo
  {volume} {110}},\ \bibinfo {pages} {124057} (\bibinfo {year} {2024})},\
  \Eprint {https://arxiv.org/abs/2409.04517} {arXiv:2409.04517 [gr-qc]}
  \BibitemShut {NoStop}%
\bibitem [{\citenamefont {Silva}\ \emph {et~al.}(2026)\citenamefont {Silva},
  \citenamefont {Tambalo}, \citenamefont {Glampedakis},\ and\ \citenamefont
  {Yagi}}]{Silva:2026jih}%
  \BibitemOpen
  \bibfield  {author} {\bibinfo {author} {\bibfnamefont {H.~O.}\ \bibnamefont
  {Silva}}, \bibinfo {author} {\bibfnamefont {G.}~\bibnamefont {Tambalo}},
  \bibinfo {author} {\bibfnamefont {K.}~\bibnamefont {Glampedakis}},\ and\
  \bibinfo {author} {\bibfnamefont {K.}~\bibnamefont {Yagi}},\ }\bibfield
  {title} {\bibinfo {title} {{Quasinormal modes and their excitation beyond
  general relativity. II. Isospectrality loss in gravitational waveforms}},\
  }\href {https://doi.org/10.1103/lgfg-lvjn} {\bibfield  {journal} {\bibinfo
  {journal} {Phys. Rev. D}\ }\textbf {\bibinfo {volume} {113}},\ \bibinfo
  {pages} {084012} (\bibinfo {year} {2026})},\ \Eprint
  {https://arxiv.org/abs/2601.13411} {arXiv:2601.13411 [gr-qc]} \BibitemShut
  {NoStop}%
\bibitem [{\citenamefont {Cano}\ and\ \citenamefont
  {David}(2025)}]{Cano:2024wzo}%
  \BibitemOpen
  \bibfield  {author} {\bibinfo {author} {\bibfnamefont {P.~A.}\ \bibnamefont
  {Cano}}\ and\ \bibinfo {author} {\bibfnamefont {M.}~\bibnamefont {David}},\
  }\bibfield  {title} {\bibinfo {title} {{Isospectrality in Effective Field
  Theory Extensions of General Relativity}},\ }\href
  {https://doi.org/10.1103/PhysRevLett.134.191401} {\bibfield  {journal}
  {\bibinfo  {journal} {Phys. Rev. Lett.}\ }\textbf {\bibinfo {volume} {134}},\
  \bibinfo {pages} {191401} (\bibinfo {year} {2025})},\ \Eprint
  {https://arxiv.org/abs/2407.12080} {arXiv:2407.12080 [hep-th]} \BibitemShut
  {NoStop}%
\bibitem [{\citenamefont {Cano}\ \emph {et~al.}(2025)\citenamefont {Cano},
  \citenamefont {David},\ and\ \citenamefont {van~der Velde}}]{Cano:2025mht}%
  \BibitemOpen
  \bibfield  {author} {\bibinfo {author} {\bibfnamefont {P.~A.}\ \bibnamefont
  {Cano}}, \bibinfo {author} {\bibfnamefont {M.}~\bibnamefont {David}},\ and\
  \bibinfo {author} {\bibfnamefont {G.}~\bibnamefont {van~der Velde}},\
  }\bibfield  {title} {\bibinfo {title} {{Eikonal quasinormal modes of
  highly-spinning black holes in higher-curvature gravity: a window into
  extremality}},\ }\href@noop {} {\  (\bibinfo {year} {2025})},\ \Eprint
  {https://arxiv.org/abs/2509.08664} {arXiv:2509.08664 [gr-qc]} \BibitemShut
  {NoStop}%
\bibitem [{\citenamefont {Gaillard}\ and\ \citenamefont
  {Zumino}(1981)}]{Gaillard:1981rj}%
  \BibitemOpen
  \bibfield  {author} {\bibinfo {author} {\bibfnamefont {M.~K.}\ \bibnamefont
  {Gaillard}}\ and\ \bibinfo {author} {\bibfnamefont {B.}~\bibnamefont
  {Zumino}},\ }\bibfield  {title} {\bibinfo {title} {{Duality Rotations for
  Interacting Fields}},\ }\href {https://doi.org/10.1016/0550-3213(81)90527-7}
  {\bibfield  {journal} {\bibinfo  {journal} {Nucl. Phys. B}\ }\textbf
  {\bibinfo {volume} {193}},\ \bibinfo {pages} {221} (\bibinfo {year}
  {1981})}\BibitemShut {NoStop}%
\bibitem [{\citenamefont {Gibbons}\ and\ \citenamefont
  {Rasheed}(1995)}]{Gibbons:1995cv}%
  \BibitemOpen
  \bibfield  {author} {\bibinfo {author} {\bibfnamefont {G.~W.}\ \bibnamefont
  {Gibbons}}\ and\ \bibinfo {author} {\bibfnamefont {D.~A.}\ \bibnamefont
  {Rasheed}},\ }\bibfield  {title} {\bibinfo {title} {{Electric - magnetic
  duality rotations in nonlinear electrodynamics}},\ }\href
  {https://doi.org/10.1016/0550-3213(95)00409-L} {\bibfield  {journal}
  {\bibinfo  {journal} {Nucl. Phys. B}\ }\textbf {\bibinfo {volume} {454}},\
  \bibinfo {pages} {185} (\bibinfo {year} {1995})},\ \Eprint
  {https://arxiv.org/abs/hep-th/9506035} {arXiv:hep-th/9506035} \BibitemShut
  {NoStop}%
\bibitem [{\citenamefont {Kol}\ and\ \citenamefont {Yau}(2023)}]{Kol:2023yxd}%
  \BibitemOpen
  \bibfield  {author} {\bibinfo {author} {\bibfnamefont {U.}~\bibnamefont
  {Kol}}\ and\ \bibinfo {author} {\bibfnamefont {S.-T.}\ \bibnamefont {Yau}},\
  }\bibfield  {title} {\bibinfo {title} {{Duality in Gauge Theory, Gravity and
  String Theory}},\ }\href@noop {} {\  (\bibinfo {year} {2023})},\ \Eprint
  {https://arxiv.org/abs/2311.07934} {arXiv:2311.07934 [hep-th]} \BibitemShut
  {NoStop}%
\bibitem [{\citenamefont {Penrose}(1960)}]{Penrose:1960eq}%
  \BibitemOpen
  \bibfield  {author} {\bibinfo {author} {\bibfnamefont {R.}~\bibnamefont
  {Penrose}},\ }\bibfield  {title} {\bibinfo {title} {{A Spinor approach to
  general relativity}},\ }\href {https://doi.org/10.1016/0003-4916(60)90021-X}
  {\bibfield  {journal} {\bibinfo  {journal} {Annals Phys.}\ }\textbf {\bibinfo
  {volume} {10}},\ \bibinfo {pages} {171} (\bibinfo {year} {1960})}\BibitemShut
  {NoStop}%
\bibitem [{\citenamefont {Nieto}(1999)}]{Nieto:1999pn}%
  \BibitemOpen
  \bibfield  {author} {\bibinfo {author} {\bibfnamefont {J.~A.}\ \bibnamefont
  {Nieto}},\ }\bibfield  {title} {\bibinfo {title} {{S duality for linearized
  gravity}},\ }\href {https://doi.org/10.1016/S0375-9601(99)00702-1} {\bibfield
   {journal} {\bibinfo  {journal} {Phys. Lett. A}\ }\textbf {\bibinfo {volume}
  {262}},\ \bibinfo {pages} {274} (\bibinfo {year} {1999})},\ \Eprint
  {https://arxiv.org/abs/hep-th/9910049} {arXiv:hep-th/9910049} \BibitemShut
  {NoStop}%
\bibitem [{\citenamefont {Hull}(2001)}]{Hull:2001iu}%
  \BibitemOpen
  \bibfield  {author} {\bibinfo {author} {\bibfnamefont {C.~M.}\ \bibnamefont
  {Hull}},\ }\bibfield  {title} {\bibinfo {title} {{Duality in gravity and
  higher spin gauge fields}},\ }\href
  {https://doi.org/10.1088/1126-6708/2001/09/027} {\bibfield  {journal}
  {\bibinfo  {journal} {JHEP}\ }\textbf {\bibinfo {volume} {09}},\ \bibinfo
  {pages} {027}},\ \Eprint {https://arxiv.org/abs/hep-th/0107149}
  {arXiv:hep-th/0107149} \BibitemShut {NoStop}%
\bibitem [{\citenamefont {Henneaux}\ and\ \citenamefont
  {Teitelboim}(2005)}]{Henneaux:2004jw}%
  \BibitemOpen
  \bibfield  {author} {\bibinfo {author} {\bibfnamefont {M.}~\bibnamefont
  {Henneaux}}\ and\ \bibinfo {author} {\bibfnamefont {C.}~\bibnamefont
  {Teitelboim}},\ }\bibfield  {title} {\bibinfo {title} {{Duality in linearized
  gravity}},\ }\href {https://doi.org/10.1103/PhysRevD.71.024018} {\bibfield
  {journal} {\bibinfo  {journal} {Phys. Rev. D}\ }\textbf {\bibinfo {volume}
  {71}},\ \bibinfo {pages} {024018} (\bibinfo {year} {2005})},\ \Eprint
  {https://arxiv.org/abs/gr-qc/0408101} {arXiv:gr-qc/0408101} \BibitemShut
  {NoStop}%
\bibitem [{\citenamefont {Deser}\ and\ \citenamefont
  {Seminara}(2005)}]{Deser:2005sz}%
  \BibitemOpen
  \bibfield  {author} {\bibinfo {author} {\bibfnamefont {S.}~\bibnamefont
  {Deser}}\ and\ \bibinfo {author} {\bibfnamefont {D.}~\bibnamefont
  {Seminara}},\ }\bibfield  {title} {\bibinfo {title} {{Free spin 2 duality
  invariance cannot be extended to GR}},\ }\href
  {https://doi.org/10.1103/PhysRevD.71.081502} {\bibfield  {journal} {\bibinfo
  {journal} {Phys. Rev. D}\ }\textbf {\bibinfo {volume} {71}},\ \bibinfo
  {pages} {081502} (\bibinfo {year} {2005})},\ \Eprint
  {https://arxiv.org/abs/hep-th/0503030} {arXiv:hep-th/0503030} \BibitemShut
  {NoStop}%
\bibitem [{\citenamefont {Bunster}\ \emph {et~al.}(2006)\citenamefont
  {Bunster}, \citenamefont {Cnockaert}, \citenamefont {Henneaux},\ and\
  \citenamefont {Portugues}}]{Bunster:2006rt}%
  \BibitemOpen
  \bibfield  {author} {\bibinfo {author} {\bibfnamefont {C.~W.}\ \bibnamefont
  {Bunster}}, \bibinfo {author} {\bibfnamefont {S.}~\bibnamefont {Cnockaert}},
  \bibinfo {author} {\bibfnamefont {M.}~\bibnamefont {Henneaux}},\ and\
  \bibinfo {author} {\bibfnamefont {R.}~\bibnamefont {Portugues}},\ }\bibfield
  {title} {\bibinfo {title} {{Monopoles for gravitation and for higher spin
  fields}},\ }\href {https://doi.org/10.1103/PhysRevD.73.105014} {\bibfield
  {journal} {\bibinfo  {journal} {Phys. Rev. D}\ }\textbf {\bibinfo {volume}
  {73}},\ \bibinfo {pages} {105014} (\bibinfo {year} {2006})},\ \Eprint
  {https://arxiv.org/abs/hep-th/0601222} {arXiv:hep-th/0601222} \BibitemShut
  {NoStop}%
\bibitem [{\citenamefont {Argurio}\ and\ \citenamefont
  {Dehouck}(2010)}]{Argurio:2009xr}%
  \BibitemOpen
  \bibfield  {author} {\bibinfo {author} {\bibfnamefont {R.}~\bibnamefont
  {Argurio}}\ and\ \bibinfo {author} {\bibfnamefont {F.}~\bibnamefont
  {Dehouck}},\ }\bibfield  {title} {\bibinfo {title} {{Gravitational duality
  and rotating solutions}},\ }\href
  {https://doi.org/10.1103/PhysRevD.81.064010} {\bibfield  {journal} {\bibinfo
  {journal} {Phys. Rev. D}\ }\textbf {\bibinfo {volume} {81}},\ \bibinfo
  {pages} {064010} (\bibinfo {year} {2010})},\ \Eprint
  {https://arxiv.org/abs/0909.0542} {arXiv:0909.0542 [hep-th]} \BibitemShut
  {NoStop}%
\bibitem [{\citenamefont {Barnich}\ and\ \citenamefont
  {Troessaert}(2009)}]{Barnich:2008ts}%
  \BibitemOpen
  \bibfield  {author} {\bibinfo {author} {\bibfnamefont {G.}~\bibnamefont
  {Barnich}}\ and\ \bibinfo {author} {\bibfnamefont {C.}~\bibnamefont
  {Troessaert}},\ }\bibfield  {title} {\bibinfo {title} {{Manifest spin 2
  duality with electric and magnetic sources}},\ }\href
  {https://doi.org/10.1088/1126-6708/2009/01/030} {\bibfield  {journal}
  {\bibinfo  {journal} {JHEP}\ }\textbf {\bibinfo {volume} {01}},\ \bibinfo
  {pages} {030}},\ \Eprint {https://arxiv.org/abs/0812.0552} {arXiv:0812.0552
  [hep-th]} \BibitemShut {NoStop}%
\bibitem [{\citenamefont {Boos}\ and\ \citenamefont
  {Kol{\'a}{\v{r}}}(2021)}]{Boos:2021suz}%
  \BibitemOpen
  \bibfield  {author} {\bibinfo {author} {\bibfnamefont {J.}~\bibnamefont
  {Boos}}\ and\ \bibinfo {author} {\bibfnamefont {I.}~\bibnamefont
  {Kol{\'a}{\v{r}}}},\ }\bibfield  {title} {\bibinfo {title} {{Nonlocality and
  gravitoelectromagnetic duality}},\ }\href
  {https://doi.org/10.1103/PhysRevD.104.024018} {\bibfield  {journal} {\bibinfo
   {journal} {Phys. Rev. D}\ }\textbf {\bibinfo {volume} {104}},\ \bibinfo
  {pages} {024018} (\bibinfo {year} {2021})},\ \Eprint
  {https://arxiv.org/abs/2103.10555} {arXiv:2103.10555 [gr-qc]} \BibitemShut
  {NoStop}%
\bibitem [{\citenamefont {Monteiro}(2024)}]{Monteiro:2023dev}%
  \BibitemOpen
  \bibfield  {author} {\bibinfo {author} {\bibfnamefont {R.}~\bibnamefont
  {Monteiro}},\ }\bibfield  {title} {\bibinfo {title} {{No U(1)
  {\textquoteleft}electric-magnetic{\textquoteright} duality in Einstein
  gravity}},\ }\href {https://doi.org/10.1007/JHEP04(2024)093} {\bibfield
  {journal} {\bibinfo  {journal} {JHEP}\ }\textbf {\bibinfo {volume} {04}},\
  \bibinfo {pages} {093}},\ \Eprint {https://arxiv.org/abs/2312.02351}
  {arXiv:2312.02351 [hep-th]} \BibitemShut {NoStop}%
\bibitem [{\citenamefont {del Rio}\ \emph {et~al.}(2025)\citenamefont {del
  Rio}, \citenamefont {Olmedo},\ and\ \citenamefont
  {Torres~Manso}}]{delRio:2025ynh}%
  \BibitemOpen
  \bibfield  {author} {\bibinfo {author} {\bibfnamefont {A.}~\bibnamefont {del
  Rio}}, \bibinfo {author} {\bibfnamefont {J.}~\bibnamefont {Olmedo}},\ and\
  \bibinfo {author} {\bibfnamefont {A.}~\bibnamefont {Torres~Manso}},\
  }\bibfield  {title} {\bibinfo {title} {{Duality symmetry and anomaly for
  gravitational waves in curved spacetimes}},\ }\href
  {https://doi.org/10.1103/th76-hy1r} {\bibfield  {journal} {\bibinfo
  {journal} {Phys. Rev. D}\ }\textbf {\bibinfo {volume} {112}},\ \bibinfo
  {pages} {085023} (\bibinfo {year} {2025})},\ \Eprint
  {https://arxiv.org/abs/2507.22588} {arXiv:2507.22588 [gr-qc]} \BibitemShut
  {NoStop}%
\bibitem [{\citenamefont {{Goebel}}(1972)}]{1972ApJ...172L..95G}%
  \BibitemOpen
  \bibfield  {author} {\bibinfo {author} {\bibfnamefont {C.~J.}\ \bibnamefont
  {{Goebel}}},\ }\bibfield  {title} {\bibinfo {title} {{Comments on the
  ``vibrations'' of a Black Hole.}},\ }\href {https://doi.org/10.1086/180898}
  {\bibfield  {journal} {\bibinfo  {journal} {Astrophys. J. Lett.}\ }\textbf
  {\bibinfo {volume} {172}},\ \bibinfo {pages} {L95} (\bibinfo {year}
  {1972})}\BibitemShut {NoStop}%
\bibitem [{\citenamefont {Ferrari}\ and\ \citenamefont
  {Mashhoon}(1984)}]{Ferrari:1984zz}%
  \BibitemOpen
  \bibfield  {author} {\bibinfo {author} {\bibfnamefont {V.}~\bibnamefont
  {Ferrari}}\ and\ \bibinfo {author} {\bibfnamefont {B.}~\bibnamefont
  {Mashhoon}},\ }\bibfield  {title} {\bibinfo {title} {{New approach to the
  quasinormal modes of a black hole}},\ }\href
  {https://doi.org/10.1103/PhysRevD.30.295} {\bibfield  {journal} {\bibinfo
  {journal} {Phys. Rev. D}\ }\textbf {\bibinfo {volume} {30}},\ \bibinfo
  {pages} {295} (\bibinfo {year} {1984})}\BibitemShut {NoStop}%
\bibitem [{\citenamefont {Mashhoon}(1985)}]{Mashhoon:1985cya}%
  \BibitemOpen
  \bibfield  {author} {\bibinfo {author} {\bibfnamefont {B.}~\bibnamefont
  {Mashhoon}},\ }\bibfield  {title} {\bibinfo {title} {{Stability of charged
  rotating black holes in the eikonal approximation}},\ }\href
  {https://doi.org/10.1103/PhysRevD.31.290} {\bibfield  {journal} {\bibinfo
  {journal} {Phys. Rev. D}\ }\textbf {\bibinfo {volume} {31}},\ \bibinfo
  {pages} {290} (\bibinfo {year} {1985})}\BibitemShut {NoStop}%
\bibitem [{\citenamefont {Berti}\ and\ \citenamefont
  {Kokkotas}(2005)}]{Berti:2005eb}%
  \BibitemOpen
  \bibfield  {author} {\bibinfo {author} {\bibfnamefont {E.}~\bibnamefont
  {Berti}}\ and\ \bibinfo {author} {\bibfnamefont {K.~D.}\ \bibnamefont
  {Kokkotas}},\ }\bibfield  {title} {\bibinfo {title} {{Quasinormal modes of
  Kerr-Newman black holes: Coupling of electromagnetic and gravitational
  perturbations}},\ }\href {https://doi.org/10.1103/PhysRevD.71.124008}
  {\bibfield  {journal} {\bibinfo  {journal} {Phys. Rev. D}\ }\textbf {\bibinfo
  {volume} {71}},\ \bibinfo {pages} {124008} (\bibinfo {year} {2005})},\
  \Eprint {https://arxiv.org/abs/gr-qc/0502065} {arXiv:gr-qc/0502065}
  \BibitemShut {NoStop}%
\bibitem [{\citenamefont {Cardoso}\ \emph {et~al.}(2009)\citenamefont
  {Cardoso}, \citenamefont {Miranda}, \citenamefont {Berti}, \citenamefont
  {Witek},\ and\ \citenamefont {Zanchin}}]{Cardoso:2008bp}%
  \BibitemOpen
  \bibfield  {author} {\bibinfo {author} {\bibfnamefont {V.}~\bibnamefont
  {Cardoso}}, \bibinfo {author} {\bibfnamefont {A.~S.}\ \bibnamefont
  {Miranda}}, \bibinfo {author} {\bibfnamefont {E.}~\bibnamefont {Berti}},
  \bibinfo {author} {\bibfnamefont {H.}~\bibnamefont {Witek}},\ and\ \bibinfo
  {author} {\bibfnamefont {V.~T.}\ \bibnamefont {Zanchin}},\ }\bibfield
  {title} {\bibinfo {title} {{Geodesic stability, Lyapunov exponents and
  quasinormal modes}},\ }\href {https://doi.org/10.1103/PhysRevD.79.064016}
  {\bibfield  {journal} {\bibinfo  {journal} {Phys. Rev. D}\ }\textbf {\bibinfo
  {volume} {79}},\ \bibinfo {pages} {064016} (\bibinfo {year} {2009})},\
  \Eprint {https://arxiv.org/abs/0812.1806} {arXiv:0812.1806 [hep-th]}
  \BibitemShut {NoStop}%
\bibitem [{\citenamefont {Dolan}(2010)}]{Dolan:2010wr}%
  \BibitemOpen
  \bibfield  {author} {\bibinfo {author} {\bibfnamefont {S.~R.}\ \bibnamefont
  {Dolan}},\ }\bibfield  {title} {\bibinfo {title} {{The Quasinormal Mode
  Spectrum of a Kerr Black Hole in the Eikonal Limit}},\ }\href
  {https://doi.org/10.1103/PhysRevD.82.104003} {\bibfield  {journal} {\bibinfo
  {journal} {Phys. Rev. D}\ }\textbf {\bibinfo {volume} {82}},\ \bibinfo
  {pages} {104003} (\bibinfo {year} {2010})},\ \Eprint
  {https://arxiv.org/abs/1007.5097} {arXiv:1007.5097 [gr-qc]} \BibitemShut
  {NoStop}%
\bibitem [{\citenamefont {Penrose}(1976)}]{Penrose1976}%
  \BibitemOpen
  \bibfield  {author} {\bibinfo {author} {\bibfnamefont {R.}~\bibnamefont
  {Penrose}},\ }\bibinfo {title} {Any space-time has a plane wave as a limit},\
  in\ \href {https://doi.org/10.1007/978-94-010-1508-0_23} {\emph {\bibinfo
  {booktitle} {Differential Geometry and Relativity: A Volume in Honour of
  Andr{\'e} Lichnerowicz on His 60th Birthday}}},\ \bibinfo {editor} {edited
  by\ \bibinfo {editor} {\bibfnamefont {M.}~\bibnamefont {Cahen}}\ and\
  \bibinfo {editor} {\bibfnamefont {M.}~\bibnamefont {Flato}}}\ (\bibinfo
  {publisher} {Springer Netherlands},\ \bibinfo {address} {Dordrecht},\
  \bibinfo {year} {1976})\ pp.\ \bibinfo {pages} {271--275}\BibitemShut
  {NoStop}%
\bibitem [{\citenamefont {Fransen}(2023)}]{Fransen:2023eqj}%
  \BibitemOpen
  \bibfield  {author} {\bibinfo {author} {\bibfnamefont {K.}~\bibnamefont
  {Fransen}},\ }\bibfield  {title} {\bibinfo {title} {{Quasinormal modes from
  Penrose limits}},\ }\href {https://doi.org/10.1088/1361-6382/acf26d}
  {\bibfield  {journal} {\bibinfo  {journal} {Class. Quant. Grav.}\ }\textbf
  {\bibinfo {volume} {40}},\ \bibinfo {pages} {205004} (\bibinfo {year}
  {2023})},\ \Eprint {https://arxiv.org/abs/2301.06999} {arXiv:2301.06999
  [gr-qc]} \BibitemShut {NoStop}%
\bibitem [{\citenamefont {Kapec}\ and\ \citenamefont
  {Sheta}(2025)}]{Kapec:2024lnr}%
  \BibitemOpen
  \bibfield  {author} {\bibinfo {author} {\bibfnamefont {D.}~\bibnamefont
  {Kapec}}\ and\ \bibinfo {author} {\bibfnamefont {A.}~\bibnamefont {Sheta}},\
  }\bibfield  {title} {\bibinfo {title} {{pp-waves and the hidden symmetries of
  black hole quasinormal modes}},\ }\href
  {https://doi.org/10.1088/1361-6382/adecda} {\bibfield  {journal} {\bibinfo
  {journal} {Class. Quant. Grav.}\ }\textbf {\bibinfo {volume} {42}},\ \bibinfo
  {pages} {155002} (\bibinfo {year} {2025})},\ \Eprint
  {https://arxiv.org/abs/2412.08551} {arXiv:2412.08551 [hep-th]} \BibitemShut
  {NoStop}%
\bibitem [{\citenamefont {Fransen}\ \emph {et~al.}(2025)\citenamefont
  {Fransen}, \citenamefont {Pere{\~n}iguez},\ and\ \citenamefont
  {Redondo-Yuste}}]{Fransen:2025cgv}%
  \BibitemOpen
  \bibfield  {author} {\bibinfo {author} {\bibfnamefont {K.}~\bibnamefont
  {Fransen}}, \bibinfo {author} {\bibfnamefont {D.}~\bibnamefont
  {Pere{\~n}iguez}},\ and\ \bibinfo {author} {\bibfnamefont {J.}~\bibnamefont
  {Redondo-Yuste}},\ }\bibfield  {title} {\bibinfo {title} {{Perturbations of
  plane waves and quadratic quasinormal modes on the lightring}},\ }\href
  {https://doi.org/10.1007/JHEP12(2025)148} {\bibfield  {journal} {\bibinfo
  {journal} {JHEP}\ }\textbf {\bibinfo {volume} {12}},\ \bibinfo {pages}
  {148}},\ \Eprint {https://arxiv.org/abs/2509.03598} {arXiv:2509.03598
  [gr-qc]} \BibitemShut {NoStop}%
\bibitem [{\citenamefont {Brinkmann}(1925)}]{Brinkmann:1925fr}%
  \BibitemOpen
  \bibfield  {author} {\bibinfo {author} {\bibfnamefont {H.~W.}\ \bibnamefont
  {Brinkmann}},\ }\bibfield  {title} {\bibinfo {title} {{Einstein spaces which
  are mapped conformally on each other}},\ }\href
  {https://doi.org/10.1007/BF01208647} {\bibfield  {journal} {\bibinfo
  {journal} {Math. Ann.}\ }\textbf {\bibinfo {volume} {94}},\ \bibinfo {pages}
  {119} (\bibinfo {year} {1925})}\BibitemShut {NoStop}%
\bibitem [{\citenamefont {Kehagias}\ \emph {et~al.}(2025)\citenamefont
  {Kehagias}, \citenamefont {Perrone},\ and\ \citenamefont
  {Riotto}}]{Kehagias:2025ntm}%
  \BibitemOpen
  \bibfield  {author} {\bibinfo {author} {\bibfnamefont {A.}~\bibnamefont
  {Kehagias}}, \bibinfo {author} {\bibfnamefont {D.}~\bibnamefont {Perrone}},\
  and\ \bibinfo {author} {\bibfnamefont {A.}~\bibnamefont {Riotto}},\
  }\bibfield  {title} {\bibinfo {title} {{Non-linear Quasi-Normal Modes of the
  Schwarzschild Black Hole from the Penrose Limit}},\ }\href@noop {} {\
  (\bibinfo {year} {2025})},\ \Eprint {https://arxiv.org/abs/2503.09350}
  {arXiv:2503.09350 [gr-qc]} \BibitemShut {NoStop}%
\bibitem [{\citenamefont {Perrone}\ \emph {et~al.}(2025)\citenamefont
  {Perrone}, \citenamefont {Kehagias},\ and\ \citenamefont
  {Riotto}}]{Perrone:2025zhy}%
  \BibitemOpen
  \bibfield  {author} {\bibinfo {author} {\bibfnamefont {D.}~\bibnamefont
  {Perrone}}, \bibinfo {author} {\bibfnamefont {A.}~\bibnamefont {Kehagias}},\
  and\ \bibinfo {author} {\bibfnamefont {A.}~\bibnamefont {Riotto}},\
  }\bibfield  {title} {\bibinfo {title} {{Nonlinearities in Kerr black hole
  ringdown from the Penrose limit}},\ }\href
  {https://doi.org/10.1088/1475-7516/2025/10/024} {\bibfield  {journal}
  {\bibinfo  {journal} {JCAP}\ }\textbf {\bibinfo {volume} {10}},\ \bibinfo
  {pages} {024}},\ \Eprint {https://arxiv.org/abs/2507.01919} {arXiv:2507.01919
  [gr-qc]} \BibitemShut {NoStop}%
\bibitem [{\citenamefont {Geroch}\ \emph {et~al.}(1973)\citenamefont {Geroch},
  \citenamefont {Held},\ and\ \citenamefont {Penrose}}]{Geroch:1973am}%
  \BibitemOpen
  \bibfield  {author} {\bibinfo {author} {\bibfnamefont {R.~P.}\ \bibnamefont
  {Geroch}}, \bibinfo {author} {\bibfnamefont {A.}~\bibnamefont {Held}},\ and\
  \bibinfo {author} {\bibfnamefont {R.}~\bibnamefont {Penrose}},\ }\bibfield
  {title} {\bibinfo {title} {{A space-time calculus based on pairs of null
  directions}},\ }\href {https://doi.org/10.1063/1.1666410} {\bibfield
  {journal} {\bibinfo  {journal} {J. Math. Phys.}\ }\textbf {\bibinfo {volume}
  {14}},\ \bibinfo {pages} {874} (\bibinfo {year} {1973})}\BibitemShut
  {NoStop}%
\bibitem [{\citenamefont {Stewart}\ and\ \citenamefont
  {Walker}(1974)}]{Stewart:1974uz}%
  \BibitemOpen
  \bibfield  {author} {\bibinfo {author} {\bibfnamefont {J.~M.}\ \bibnamefont
  {Stewart}}\ and\ \bibinfo {author} {\bibfnamefont {M.}~\bibnamefont
  {Walker}},\ }\bibfield  {title} {\bibinfo {title} {{Perturbations of
  spacetimes in general relativity}},\ }\href
  {https://doi.org/10.1098/rspa.1974.0172} {\bibfield  {journal} {\bibinfo
  {journal} {Proc. Roy. Soc. Lond. A}\ }\textbf {\bibinfo {volume} {341}},\
  \bibinfo {pages} {49} (\bibinfo {year} {1974})}\BibitemShut {NoStop}%
\bibitem [{\citenamefont {Green}\ \emph {et~al.}(2023)\citenamefont {Green},
  \citenamefont {Hollands}, \citenamefont {Sberna}, \citenamefont {Toomani},\
  and\ \citenamefont {Zimmerman}}]{Green:2022htq}%
  \BibitemOpen
  \bibfield  {author} {\bibinfo {author} {\bibfnamefont {S.~R.}\ \bibnamefont
  {Green}}, \bibinfo {author} {\bibfnamefont {S.}~\bibnamefont {Hollands}},
  \bibinfo {author} {\bibfnamefont {L.}~\bibnamefont {Sberna}}, \bibinfo
  {author} {\bibfnamefont {V.}~\bibnamefont {Toomani}},\ and\ \bibinfo {author}
  {\bibfnamefont {P.}~\bibnamefont {Zimmerman}},\ }\bibfield  {title} {\bibinfo
  {title} {{Conserved currents for a Kerr black hole and orthogonality of
  quasinormal modes}},\ }\href {https://doi.org/10.1103/PhysRevD.107.064030}
  {\bibfield  {journal} {\bibinfo  {journal} {Phys. Rev. D}\ }\textbf {\bibinfo
  {volume} {107}},\ \bibinfo {pages} {064030} (\bibinfo {year} {2023})},\
  \Eprint {https://arxiv.org/abs/2210.15935} {arXiv:2210.15935 [gr-qc]}
  \BibitemShut {NoStop}%
\bibitem [{\citenamefont {Aksteiner}\ and\ \citenamefont
  {Andersson}(2011)}]{Aksteiner:2010rh}%
  \BibitemOpen
  \bibfield  {author} {\bibinfo {author} {\bibfnamefont {S.}~\bibnamefont
  {Aksteiner}}\ and\ \bibinfo {author} {\bibfnamefont {L.}~\bibnamefont
  {Andersson}},\ }\bibfield  {title} {\bibinfo {title} {{Linearized gravity and
  gauge conditions}},\ }\href {https://doi.org/10.1088/0264-9381/28/6/065001}
  {\bibfield  {journal} {\bibinfo  {journal} {Class. Quant. Grav.}\ }\textbf
  {\bibinfo {volume} {28}},\ \bibinfo {pages} {065001} (\bibinfo {year}
  {2011})},\ \Eprint {https://arxiv.org/abs/1009.5647} {arXiv:1009.5647
  [gr-qc]} \BibitemShut {NoStop}%
\bibitem [{\citenamefont {Aksteiner}\ \emph {et~al.}(2019)\citenamefont
  {Aksteiner}, \citenamefont {Andersson},\ and\ \citenamefont
  {B{\"a}ckdahl}}]{Aksteiner:2016pjt}%
  \BibitemOpen
  \bibfield  {author} {\bibinfo {author} {\bibfnamefont {S.}~\bibnamefont
  {Aksteiner}}, \bibinfo {author} {\bibfnamefont {L.}~\bibnamefont
  {Andersson}},\ and\ \bibinfo {author} {\bibfnamefont {T.}~\bibnamefont
  {B{\"a}ckdahl}},\ }\bibfield  {title} {\bibinfo {title} {{New identities for
  linearized gravity on the Kerr spacetime}},\ }\href
  {https://doi.org/10.1103/PhysRevD.99.044043} {\bibfield  {journal} {\bibinfo
  {journal} {Phys. Rev. D}\ }\textbf {\bibinfo {volume} {99}},\ \bibinfo
  {pages} {044043} (\bibinfo {year} {2019})},\ \Eprint
  {https://arxiv.org/abs/1601.06084} {arXiv:1601.06084 [gr-qc]} \BibitemShut
  {NoStop}%
\bibitem [{\citenamefont {{Calkin}}(1965)}]{1965AmJPh..33..958C}%
  \BibitemOpen
  \bibfield  {author} {\bibinfo {author} {\bibfnamefont {M.~G.}\ \bibnamefont
  {{Calkin}}},\ }\bibfield  {title} {\bibinfo {title} {{An Invariance Property
  of the Free Electromagnetic Field}},\ }\href
  {https://doi.org/10.1119/1.1971089} {\bibfield  {journal} {\bibinfo
  {journal} {American Journal of Physics}\ }\textbf {\bibinfo {volume} {33}},\
  \bibinfo {pages} {958} (\bibinfo {year} {1965})}\BibitemShut {NoStop}%
\bibitem [{\citenamefont {Ananth}\ \emph {et~al.}(2006)\citenamefont {Ananth},
  \citenamefont {Brink}, \citenamefont {Heise},\ and\ \citenamefont
  {Svendsen}}]{Ananth:2006fh}%
  \BibitemOpen
  \bibfield  {author} {\bibinfo {author} {\bibfnamefont {S.}~\bibnamefont
  {Ananth}}, \bibinfo {author} {\bibfnamefont {L.}~\bibnamefont {Brink}},
  \bibinfo {author} {\bibfnamefont {R.}~\bibnamefont {Heise}},\ and\ \bibinfo
  {author} {\bibfnamefont {H.~G.}\ \bibnamefont {Svendsen}},\ }\bibfield
  {title} {\bibinfo {title} {{The N=8 Supergravity Hamiltonian as a Quadratic
  Form}},\ }\href {https://doi.org/10.1016/j.nuclphysb.2006.07.014} {\bibfield
  {journal} {\bibinfo  {journal} {Nucl. Phys. B}\ }\textbf {\bibinfo {volume}
  {753}},\ \bibinfo {pages} {195} (\bibinfo {year} {2006})},\ \Eprint
  {https://arxiv.org/abs/hep-th/0607019} {arXiv:hep-th/0607019} \BibitemShut
  {NoStop}%
\bibitem [{\citenamefont {Schwarz}(1995)}]{Schwarz:1995dk}%
  \BibitemOpen
  \bibfield  {author} {\bibinfo {author} {\bibfnamefont {J.~H.}\ \bibnamefont
  {Schwarz}},\ }\bibfield  {title} {\bibinfo {title} {{An SL(2,Z) multiplet of
  type IIB superstrings}},\ }\href
  {https://doi.org/10.1016/0370-2693(95)01405-5} {\bibfield  {journal}
  {\bibinfo  {journal} {Phys. Lett. B}\ }\textbf {\bibinfo {volume} {360}},\
  \bibinfo {pages} {13} (\bibinfo {year} {1995})},\ \bibinfo {note} {[Erratum:
  Phys.Lett.B 364, 252 (1995)]},\ \Eprint
  {https://arxiv.org/abs/hep-th/9508143} {arXiv:hep-th/9508143} \BibitemShut
  {NoStop}%
\bibitem [{\citenamefont {Wald}(1978)}]{Wald:1978vm}%
  \BibitemOpen
  \bibfield  {author} {\bibinfo {author} {\bibfnamefont {R.~M.}\ \bibnamefont
  {Wald}},\ }\bibfield  {title} {\bibinfo {title} {{Construction of Solutions
  of Gravitational, Electromagnetic, Or Other Perturbation Equations from
  Solutions of Decoupled Equations}},\ }\href
  {https://doi.org/10.1103/PhysRevLett.41.203} {\bibfield  {journal} {\bibinfo
  {journal} {Phys. Rev. Lett.}\ }\textbf {\bibinfo {volume} {41}},\ \bibinfo
  {pages} {203} (\bibinfo {year} {1978})}\BibitemShut {NoStop}%
\bibitem [{\citenamefont {{Kegeles}}\ and\ \citenamefont
  {{Cohen}}(1979)}]{Kegeles}%
  \BibitemOpen
  \bibfield  {author} {\bibinfo {author} {\bibfnamefont {L.~S.}\ \bibnamefont
  {{Kegeles}}}\ and\ \bibinfo {author} {\bibfnamefont {J.~M.}\ \bibnamefont
  {{Cohen}}},\ }\bibfield  {title} {\bibinfo {title} {{Constructive procedure
  for perturbations of spacetimes}},\ }\href
  {https://doi.org/10.1103/PhysRevD.19.1641} {\bibfield  {journal} {\bibinfo
  {journal} {\prd}\ }\textbf {\bibinfo {volume} {19}},\ \bibinfo {pages} {1641}
  (\bibinfo {year} {1979})}\BibitemShut {NoStop}%
\bibitem [{\citenamefont {Hadar}\ \emph {et~al.}(2022)\citenamefont {Hadar},
  \citenamefont {Kapec}, \citenamefont {Lupsasca},\ and\ \citenamefont
  {Strominger}}]{Hadar:2022xag}%
  \BibitemOpen
  \bibfield  {author} {\bibinfo {author} {\bibfnamefont {S.}~\bibnamefont
  {Hadar}}, \bibinfo {author} {\bibfnamefont {D.}~\bibnamefont {Kapec}},
  \bibinfo {author} {\bibfnamefont {A.}~\bibnamefont {Lupsasca}},\ and\
  \bibinfo {author} {\bibfnamefont {A.}~\bibnamefont {Strominger}},\ }\bibfield
   {title} {\bibinfo {title} {{Holography of the photon ring}},\ }\href
  {https://doi.org/10.1088/1361-6382/ac8d43} {\bibfield  {journal} {\bibinfo
  {journal} {Class. Quant. Grav.}\ }\textbf {\bibinfo {volume} {39}},\ \bibinfo
  {pages} {215001} (\bibinfo {year} {2022})},\ \Eprint
  {https://arxiv.org/abs/2205.05064} {arXiv:2205.05064 [gr-qc]} \BibitemShut
  {NoStop}%
\bibitem [{\citenamefont {Despontin}\ \emph {et~al.}(2025)\citenamefont
  {Despontin}, \citenamefont {Detournay},\ and\ \citenamefont
  {Fontaine}}]{Despontin:2025svw}%
  \BibitemOpen
  \bibfield  {author} {\bibinfo {author} {\bibfnamefont {E.}~\bibnamefont
  {Despontin}}, \bibinfo {author} {\bibfnamefont {S.}~\bibnamefont
  {Detournay}},\ and\ \bibinfo {author} {\bibfnamefont {D.}~\bibnamefont
  {Fontaine}},\ }\bibfield  {title} {\bibinfo {title} {{Infinite-dimensional
  symmetries in plane wave spacetimes}}\ }\href
  {https://doi.org/10.1103/2q94-5crl} {10.1103/2q94-5crl} (\bibinfo {year}
  {2025}),\ \Eprint {https://arxiv.org/abs/2508.12760} {arXiv:2508.12760
  [hep-th]} \BibitemShut {NoStop}%
\bibitem [{\citenamefont {Li}\ \emph {et~al.}(2024)\citenamefont {Li},
  \citenamefont {Hussain}, \citenamefont {Wagle}, \citenamefont {Chen},
  \citenamefont {Yunes},\ and\ \citenamefont {Zimmerman}}]{Li:2023ulk}%
  \BibitemOpen
  \bibfield  {author} {\bibinfo {author} {\bibfnamefont {D.}~\bibnamefont
  {Li}}, \bibinfo {author} {\bibfnamefont {A.}~\bibnamefont {Hussain}},
  \bibinfo {author} {\bibfnamefont {P.}~\bibnamefont {Wagle}}, \bibinfo
  {author} {\bibfnamefont {Y.}~\bibnamefont {Chen}}, \bibinfo {author}
  {\bibfnamefont {N.}~\bibnamefont {Yunes}},\ and\ \bibinfo {author}
  {\bibfnamefont {A.}~\bibnamefont {Zimmerman}},\ }\bibfield  {title} {\bibinfo
  {title} {{Isospectrality breaking in the Teukolsky formalism}},\ }\href
  {https://doi.org/10.1103/PhysRevD.109.104026} {\bibfield  {journal} {\bibinfo
   {journal} {Phys. Rev. D}\ }\textbf {\bibinfo {volume} {109}},\ \bibinfo
  {pages} {104026} (\bibinfo {year} {2024})},\ \Eprint
  {https://arxiv.org/abs/2310.06033} {arXiv:2310.06033 [gr-qc]} \BibitemShut
  {NoStop}%
\bibitem [{\citenamefont {Endlich}\ \emph {et~al.}(2017)\citenamefont
  {Endlich}, \citenamefont {Gorbenko}, \citenamefont {Huang},\ and\
  \citenamefont {Senatore}}]{Endlich:2017tqa}%
  \BibitemOpen
  \bibfield  {author} {\bibinfo {author} {\bibfnamefont {S.}~\bibnamefont
  {Endlich}}, \bibinfo {author} {\bibfnamefont {V.}~\bibnamefont {Gorbenko}},
  \bibinfo {author} {\bibfnamefont {J.}~\bibnamefont {Huang}},\ and\ \bibinfo
  {author} {\bibfnamefont {L.}~\bibnamefont {Senatore}},\ }\bibfield  {title}
  {\bibinfo {title} {{An effective formalism for testing extensions to General
  Relativity with gravitational waves}},\ }\href
  {https://doi.org/10.1007/JHEP09(2017)122} {\bibfield  {journal} {\bibinfo
  {journal} {JHEP}\ }\textbf {\bibinfo {volume} {09}},\ \bibinfo {pages}
  {122}},\ \Eprint {https://arxiv.org/abs/1704.01590} {arXiv:1704.01590
  [gr-qc]} \BibitemShut {NoStop}%
\bibitem [{\citenamefont {Cano}\ and\ \citenamefont
  {Ruip{\'e}rez}(2019)}]{Cano:2019ore}%
  \BibitemOpen
  \bibfield  {author} {\bibinfo {author} {\bibfnamefont {P.~A.}\ \bibnamefont
  {Cano}}\ and\ \bibinfo {author} {\bibfnamefont {A.}~\bibnamefont
  {Ruip{\'e}rez}},\ }\bibfield  {title} {\bibinfo {title} {{Leading
  higher-derivative corrections to Kerr geometry}},\ }\href
  {https://doi.org/10.1007/JHEP05(2019)189} {\bibfield  {journal} {\bibinfo
  {journal} {JHEP}\ }\textbf {\bibinfo {volume} {05}},\ \bibinfo {pages}
  {189}},\ \bibinfo {note} {[Erratum: JHEP 03, 187 (2020)]},\ \Eprint
  {https://arxiv.org/abs/1901.01315} {arXiv:1901.01315 [gr-qc]} \BibitemShut
  {NoStop}%
\bibitem [{\citenamefont {Gruzinov}\ and\ \citenamefont
  {Kleban}(2007)}]{Gruzinov:2006ie}%
  \BibitemOpen
  \bibfield  {author} {\bibinfo {author} {\bibfnamefont {A.}~\bibnamefont
  {Gruzinov}}\ and\ \bibinfo {author} {\bibfnamefont {M.}~\bibnamefont
  {Kleban}},\ }\bibfield  {title} {\bibinfo {title} {{Causality Constrains
  Higher Curvature Corrections to Gravity}},\ }\href
  {https://doi.org/10.1088/0264-9381/24/13/N02} {\bibfield  {journal} {\bibinfo
   {journal} {Class. Quant. Grav.}\ }\textbf {\bibinfo {volume} {24}},\
  \bibinfo {pages} {3521} (\bibinfo {year} {2007})},\ \Eprint
  {https://arxiv.org/abs/hep-th/0612015} {arXiv:hep-th/0612015} \BibitemShut
  {NoStop}%
\bibitem [{\citenamefont {Okounkova}\ \emph {et~al.}(2022)\citenamefont
  {Okounkova}, \citenamefont {Farr}, \citenamefont {Isi},\ and\ \citenamefont
  {Stein}}]{Okounkova:2021xjv}%
  \BibitemOpen
  \bibfield  {author} {\bibinfo {author} {\bibfnamefont {M.}~\bibnamefont
  {Okounkova}}, \bibinfo {author} {\bibfnamefont {W.~M.}\ \bibnamefont {Farr}},
  \bibinfo {author} {\bibfnamefont {M.}~\bibnamefont {Isi}},\ and\ \bibinfo
  {author} {\bibfnamefont {L.~C.}\ \bibnamefont {Stein}},\ }\bibfield  {title}
  {\bibinfo {title} {{Constraining gravitational wave amplitude birefringence
  and Chern-Simons gravity with GWTC-2}},\ }\href
  {https://doi.org/10.1103/PhysRevD.106.044067} {\bibfield  {journal} {\bibinfo
   {journal} {Phys. Rev. D}\ }\textbf {\bibinfo {volume} {106}},\ \bibinfo
  {pages} {044067} (\bibinfo {year} {2022})},\ \Eprint
  {https://arxiv.org/abs/2101.11153} {arXiv:2101.11153 [gr-qc]} \BibitemShut
  {NoStop}%
\bibitem [{\citenamefont {Novotn{\'y}}(2018)}]{Novotny:2018iph}%
  \BibitemOpen
  \bibfield  {author} {\bibinfo {author} {\bibfnamefont {J.}~\bibnamefont
  {Novotn{\'y}}},\ }\bibfield  {title} {\bibinfo {title} {{Self-duality,
  helicity conservation and normal ordering in nonlinear QED}},\ }\href
  {https://doi.org/10.1103/PhysRevD.98.085015} {\bibfield  {journal} {\bibinfo
  {journal} {Phys. Rev. D}\ }\textbf {\bibinfo {volume} {98}},\ \bibinfo
  {pages} {085015} (\bibinfo {year} {2018})},\ \Eprint
  {https://arxiv.org/abs/1806.02167} {arXiv:1806.02167 [hep-th]} \BibitemShut
  {NoStop}%
\bibitem [{\citenamefont {Rosly}\ and\ \citenamefont
  {Selivanov}(2002)}]{Rosly:2002jt}%
  \BibitemOpen
  \bibfield  {author} {\bibinfo {author} {\bibfnamefont {A.~A.}\ \bibnamefont
  {Rosly}}\ and\ \bibinfo {author} {\bibfnamefont {K.~G.}\ \bibnamefont
  {Selivanov}},\ }\bibfield  {title} {\bibinfo {title} {{Helicity conservation
  in Born-Infeld theory}},\ }in\ \href@noop {} {\emph {\bibinfo {booktitle}
  {{Workshop on String Theory and Complex Geometry}}}}\ (\bibinfo {year}
  {2002})\ \Eprint {https://arxiv.org/abs/hep-th/0204229}
  {arXiv:hep-th/0204229} \BibitemShut {NoStop}%
\bibitem [{\citenamefont {Pope}\ \emph {et~al.}(2025)\citenamefont {Pope},
  \citenamefont {Rohrer},\ and\ \citenamefont {Whiting}}]{Pope:2025jgz}%
  \BibitemOpen
  \bibfield  {author} {\bibinfo {author} {\bibfnamefont {C.~N.}\ \bibnamefont
  {Pope}}, \bibinfo {author} {\bibfnamefont {D.~O.}\ \bibnamefont {Rohrer}},\
  and\ \bibinfo {author} {\bibfnamefont {B.~F.}\ \bibnamefont {Whiting}},\
  }\bibfield  {title} {\bibinfo {title} {{Perturbations of black holes in
  Einstein-Maxwell-dilaton-axion theories}},\ }\href
  {https://doi.org/10.1103/1b6k-f38p} {\bibfield  {journal} {\bibinfo
  {journal} {Phys. Rev. D}\ }\textbf {\bibinfo {volume} {112}},\ \bibinfo
  {pages} {124064} (\bibinfo {year} {2025})},\ \Eprint
  {https://arxiv.org/abs/2508.04589} {arXiv:2508.04589 [hep-th]} \BibitemShut
  {NoStop}%
\bibitem [{\citenamefont {Boyce}\ and\ \citenamefont
  {Santos}(2026)}]{Boyce:2026rnn}%
  \BibitemOpen
  \bibfield  {author} {\bibinfo {author} {\bibfnamefont {W.~L.}\ \bibnamefont
  {Boyce}}\ and\ \bibinfo {author} {\bibfnamefont {J.~E.}\ \bibnamefont
  {Santos}},\ }\bibfield  {title} {\bibinfo {title} {{Kerr Black Hole Ringdown
  in Effective Field Theory}},\ }\href@noop {} {\  (\bibinfo {year} {2026})},\
  \Eprint {https://arxiv.org/abs/2603.10102} {arXiv:2603.10102 [gr-qc]}
  \BibitemShut {NoStop}%
\bibitem [{\citenamefont {Hull}(2000)}]{Hull:2000rr}%
  \BibitemOpen
  \bibfield  {author} {\bibinfo {author} {\bibfnamefont {C.~M.}\ \bibnamefont
  {Hull}},\ }\bibfield  {title} {\bibinfo {title} {{Symmetries and
  compactifications of (4,0) conformal gravity}},\ }\href
  {https://doi.org/10.1088/1126-6708/2000/12/007} {\bibfield  {journal}
  {\bibinfo  {journal} {JHEP}\ }\textbf {\bibinfo {volume} {12}},\ \bibinfo
  {pages} {007}},\ \Eprint {https://arxiv.org/abs/hep-th/0011215}
  {arXiv:hep-th/0011215} \BibitemShut {NoStop}%
\bibitem [{\citenamefont {Pereñiguez}(2025)}]{2spinorNotes}%
  \BibitemOpen
  \bibfield  {author} {\bibinfo {author} {\bibfnamefont {D.}~\bibnamefont
  {Pereñiguez}},\ }\href {https://the-center-of-gravity.com/lecture-notes/}
  {\bibinfo {title} {Lecture notes on 2-spinors in general relativity}}
  (\bibinfo {year} {2025})\BibitemShut {NoStop}%
\bibitem [{\citenamefont {Bini}\ \emph {et~al.}(2002)\citenamefont {Bini},
  \citenamefont {Cherubini}, \citenamefont {Jantzen},\ and\ \citenamefont
  {Ruffini}}]{Bini:2002jx}%
  \BibitemOpen
  \bibfield  {author} {\bibinfo {author} {\bibfnamefont {D.}~\bibnamefont
  {Bini}}, \bibinfo {author} {\bibfnamefont {C.}~\bibnamefont {Cherubini}},
  \bibinfo {author} {\bibfnamefont {R.~T.}\ \bibnamefont {Jantzen}},\ and\
  \bibinfo {author} {\bibfnamefont {R.~J.}\ \bibnamefont {Ruffini}},\
  }\bibfield  {title} {\bibinfo {title} {{Teukolsky master equation: De Rham
  wave equation for the gravitational and electromagnetic fields in vacuum}},\
  }\href {https://doi.org/10.1143/PTP.107.967} {\bibfield  {journal} {\bibinfo
  {journal} {Prog. Theor. Phys.}\ }\textbf {\bibinfo {volume} {107}},\ \bibinfo
  {pages} {967} (\bibinfo {year} {2002})},\ \Eprint
  {https://arxiv.org/abs/gr-qc/0203069} {arXiv:gr-qc/0203069} \BibitemShut
  {NoStop}%
\bibitem [{\citenamefont {Zimmerman}\ \emph {et~al.}(2015)\citenamefont
  {Zimmerman}, \citenamefont {Yang}, \citenamefont {Mark}, \citenamefont
  {Chen},\ and\ \citenamefont {Lehner}}]{Zimmerman:2014aha}%
  \BibitemOpen
  \bibfield  {author} {\bibinfo {author} {\bibfnamefont {A.}~\bibnamefont
  {Zimmerman}}, \bibinfo {author} {\bibfnamefont {H.}~\bibnamefont {Yang}},
  \bibinfo {author} {\bibfnamefont {Z.}~\bibnamefont {Mark}}, \bibinfo {author}
  {\bibfnamefont {Y.}~\bibnamefont {Chen}},\ and\ \bibinfo {author}
  {\bibfnamefont {L.}~\bibnamefont {Lehner}},\ }\bibfield  {title} {\bibinfo
  {title} {{Quasinormal Modes Beyond Kerr}},\ }\href
  {https://doi.org/10.1007/978-3-319-10488-1_19} {\bibfield  {journal}
  {\bibinfo  {journal} {Astrophys. Space Sci. Proc.}\ }\textbf {\bibinfo
  {volume} {40}},\ \bibinfo {pages} {217} (\bibinfo {year} {2015})},\ \Eprint
  {https://arxiv.org/abs/1406.4206} {arXiv:1406.4206 [gr-qc]} \BibitemShut
  {NoStop}%
\end{thebibliography}%

\clearpage

\onecolumngrid   

\section*{Supplemental Material}

\addcontentsline{toc}{section}{Supplemental Material}

\section{Curvature of Penrose-limit pp-waves}
\noindent
The Riemann tensor of \eqref{eq:PL} takes the form
\begin{equation}
\begin{aligned}\label{eq:RiemanFrame}
    R&=\left[-\frac{1}{2}\left(A_{22}(u)-A_{11}(u)\right)-i A_{12}(u)\right]\left(\ell\wedge m\right)\otimes\left(\ell\wedge m\right)+\left[\frac{1}{2}\left(A_{11}(u)+A_{22}(u)\right)\right]\left(\ell\wedge m\right)\otimes\left(\ell\wedge \bar{m}\right)\\
    &+c.c.
\end{aligned}
\end{equation}
The only nonvanishing spin coefficient and curvature scalar are
\begin{equation}
    \kappa'=\frac{1}{2\sqrt{2}}\left(\partial_{x}-i\partial_{y}\right)\left(A_{ij}(u)x^{i}x^{j}\right)\, ,\qquad \Psi_{4}=-\frac{1}{4}\left(\partial_{x}-i\partial_{y}\right)^{2}\left(A_{ij}(u)x^{i}x^{j}\right)\, .
\end{equation}

\section{$h^{\pm}$ from $\Psi^{\pm}_\text{\tiny H}$}
\noindent
To identify which Hertz potentials generate even and odd metric perturbations, we follow~\cite{Li:2023ulk} and introduce the operator $\mathcal{P}=PC$, where $C$ denotes complex conjugation. Since $C^{2}=1$, $[P,C]=[\mathcal{P},C]=0$, and $\mathcal{P}^{2}=1$, one also has
\begin{equation}\label{eq:OpProp}
\left[\mathcal{P},\tho\right]=\left[\mathcal{P},\tho'\right]=0\, , \quad  \left\{\mathcal{P},\eth\right\}=\left\{\mathcal{P},\eth'\right\}=0\,,
\end{equation}
where $[-,-]$ and $\{-,-\}$ denote the commutator and anti-commutator, respectively. Computing the even and odd parts of 
$h(\Psi_\text{\tiny H})$ in \eqref{eq:recmet2_final}, $h^{\pm}(\Psi_\text{\tiny H})=(h(\Psi_\text{\tiny H})\pm P h(\Psi_\text{\tiny H}))/2$, one finds after using \eqref{eq:OpProp} that these are given by $h^{\pm}(\Psi_\text{\tiny H})=h^{\pm}(\Psi^{\pm}_\text{\tiny H})$, with $\Psi_\text{\tiny H}^{\pm}\equiv (\Psi_\text{\tiny H}\pm\mathcal{P}\Psi_\text{\tiny H})/2$.

\section{The curvature GHP equations beyond GR}\label{app:Curvature_eqs}
\noindent
To derive the modified perturbation equations arising from \eqref{eq:HD_Eqs}, we follow the approach of \cite{Fransen:2025cgv}, employing 2-spinors, now extended to account for Ricci non-flatness. In a straightforward manner, one obtains an exact identity satisfied by the curvature spinors,
\begin{equation}
\begin{aligned}
    \square\Psi_{ABCD}-6\tensor{\Psi}{^E^F_{(AB}}\tensor{\Psi}{_{CD)}_E_F}=12 \Lambda \Psi_{ABCD}+2\nabla_{AA'}\nabla_{BB'}\tensor{\Phi}{_C_D^{A'}^{B'}}+4\nabla_{AA'}\left(\epsilon_{B(C}\tensor{\nabla}{_{D)}^{A'}}\Lambda\right)\, .
\end{aligned}
\end{equation}
This follows by acting with $\nabla^{AA'}$ on the differential Bianchi identity for $\Psi_{ABCD}$ and using the spinor Ricci identity. Here, $\Phi_{ABC'D'}$ denotes the Ricci curvature spinor, and $\Lambda=R/24$, with $R$ the Ricci scalar. Projecting this equation onto the spinor dyad components $o^{A}o^{B}o^{C}o^{D}$, $o^{A}o^{B}o^{C}\iota^{D}$, $\dots$, $\iota^{A}\iota^{B}\iota^{C}\iota^{D}$, and expanding all quantities in their GHP components, yields nonperturbative GHP identities that take the form of wave equations for $\Psi_{0}, \Psi_{1}, \ldots, \Psi_{4}$. These equations generalize those in \cite{Stewart:1974uz,Bini:2002jx} by incorporating contributions from the Ricci tensor; in fact, they are exact identities. Their explicit form, particularly the dependence on the GHP components of $\Phi_{ABC'D'}$, is lengthy and not displayed here, although it can be efficiently derived using \texttt{xAct} packages such as \texttt{SpinFrames}.

The next step is to linearize these equations for fully arbitrary perturbations around the background \eqref{eq:PLLR}. In addition to perturbations of the GHP quantities, one must consider perturbations of the metric $h_{ab}$ and of the frame vectors $\dot{\ell}^{a}$, $\dot{n}^{a}$, and $\dot{m}^{a}$ (we define frame perturbations with indices up, so that, for example, $\left(m_{a}\right)\dot{}=h_{ab}m^{b}+\dot{m}_{a}$). As discussed in \cite{Fransen:2025cgv}, one can adopt the geodesic, parallel, and transverse (GPT) gauge, defined by
\begin{equation}\label{eq:geod_Trans}
    \dot{\ell}^{a}=0\, , \quad \ell^{a}h_{ab}=0\, ,
\end{equation}
with $\dot{n}^{a}$ and $\dot{m}^{a}$ given by
\begin{equation}\label{frame2}
\begin{aligned}
    \dot{n}^{a}&=\frac{1}{2}\left(-h_{nn}\ell^{a}+h_{n\bar{m}}m^{a}+h_{nm}\bar{m}^{a}\right)\, ,\\
    \dot{m}^{a}&=\frac{1}{2}\left(-h_{nm}\ell^{a}+h_{m\bar{m}}m^{a}+h_{m m}\bar{m}^{a}\right)\, ,
\end{aligned}    
\end{equation}
so that the frame satisfies the parallel transport condition
\begin{equation}\label{eq:parallelcond}
\left(\ell^{b}\nabla_{b}\ell^{a}\right)\dot{}=\left(\ell^{b}\nabla_{b}n^{a}\right)\dot{}=\left(\ell^{b}\nabla_{b}m^{a}\right)\dot{}=0\, ,
\end{equation}
and also $\left(n_{a}\right)\dot{}=-\dot{n}_{a}$ and $\left(m_{a}\right)\dot{}=-\dot{m}_{a}$. A key advantage of this gauge is that the variations of the following spin coefficients vanish:
\begin{equation}\label{eq:zeroSC}
\dot{\kappa}=\dot{\tau}'=\dot{\varepsilon}=0\,.
\end{equation}
The residual gauge freedom compatible with the GPT conditions \eqref{eq:geod_Trans} and \eqref{eq:parallelcond} is detailed in \cite{Fransen:2025cgv}. Given the simplicity of the background and this gauge choice, the linearized curvature equations simplify significantly, yielding
\begin{align}\label{eq:Psi0CWE}\notag
    2 \tho' \tho \dot{\Psi}_{0}{} - 2 \eth' \eth \dot{\Psi}_{0}{}&=2 \tho \tho \dot{\Phi}_{02'}{} - 4 \eth \tho \dot{\Phi}_{01'}{} + 2 \
\eth \eth \dot{\Phi}_{00'}{}\\
&\equiv S_{0}\, ,\\ \notag \\\notag
 2 \thop \tho \dot{\Psi}_{1}{} - 2 \
\eth' \eth \dot{\Psi}_{1}{}+2 \kappa ' \tho \dot{\Psi}_{0}{}&=2 \tho \tho \dot{\Phi}_{12'}{} - 4 \eth \tho \dot{\Phi}_{11'}{} + 2 \
\eth \eth \dot{\Phi}_{10'}{}\\
&\equiv S_{1}\, ,\\ \notag \\\notag
 2 \thop \tho \dot{\Psi}_{2}{} - 2 \eth' \eth \dot{\Psi}_{2}{}-2 \dot{\Psi}_{0}{} \Psi_{4} + 4 \kappa ' \tho \dot{\Psi}_{1}{} &=2 \tho \tho \dot{\Phi}_{22'}{} + 4 \thop \tho \dot{\Lambda} - 4 \eth \
\tho \dot{\Phi}_{21'}{} + 2 \eth \eth \dot{\Phi}_{20'}{} - 4 \eth' \
\eth \dot{\Lambda} \\
&\equiv S_{2}\, ,\\ \notag \\\notag
 2 \thop \tho \dot{\Psi}_{3}{}- 2 \
\eth' \eth \dot{\Psi}_{3}{}-4 \dot{\Psi}_{1}{} \Psi_{4} + 6 \
\kappa ' \eth' \dot{\Psi}_{1}{} - 2 \Psi_{4} \eth' \dot{\sigma} &=6 \kappa ' \tho \dot{\Phi}_{11'}{} + 6 \kappa ' \tho \dot{\Lambda} + \
4 \kappa ' \thop \dot{\Phi}_{00'}{} + 2 \thop \thop \
\dot{\Phi}_{10'}{} \\ \notag
&- 4 \kappa ' \eth \dot{\Phi}_{10'}{} - 4 \kappa ' \
\eth' \dot{\Phi}_{01'}{} - 2 \bar{\kappa}' \eth' \dot{\Phi}_{10'}{} \\ \notag 
&- 2 \dot{\Phi}_{01'}{} \eth' \kappa '  - 4 \eth' \thop \dot{\Phi}_{11'}{} + 2 \eth' \eth' \
\dot{\Phi}_{12'}{} \\
&\equiv S_{3}\, ,\\ \notag \\\notag
2 \thop \tho \dot{\Psi}_{4}{}-4 \dot{\Psi}_{2}{} \Psi_{4} + 2 \
\kappa ' \eth' \dot{\Psi}_{2}{} - 2 \Psi_{4} \eth' \dot{\tau} - 6 \
\dot{\Psi}_{2}{} \eth' \kappa ' - 2 \eth' \thop \dot{\Psi}_{3}{}&=4 \dot{\Phi}_{11'}{} \Psi_{4} + 4 \dot{\Lambda} \Psi_{4} - 4 \
\dot{\Phi}_{00'}{} \kappa '^2 + 2 \thop \thop \dot{\Phi}_{20'}{} \\ \notag 
&+ 2 \
\kappa ' \eth \dot{\Phi}_{20'}{} + 4 \kappa ' \eth' \
\dot{\Phi}_{11'}{} - 8 \kappa ' \eth' \dot{\Lambda} + 4 \
\dot{\Phi}_{11'}{} \eth' \kappa ' - 2 \eth' \thop \dot{\Phi}_{21'}{}\\ \label{eq:Psi4CWE}
&\equiv S_{4}\, .
\end{align}
We recall that, on the background \eqref{eq:PLLR}, the wave operator takes the form
\begin{equation}
    \Box = \nabla_{a}\nabla^{a}=2\left(\thop\tho-\eth'\eth\right)\, ,
\end{equation}
so these equations indeed describe wave propagation for the curvature scalars. In deriving them, we also employed the linearized Bianchi identity
\begin{equation}
    \dot{\eth}\Psi_{4}=\dot{\tau} \Psi_{4} + 2 \dot{\Phi}_{11'}{} \kappa ' + 3 \
\dot{\Psi}_{2}{} \kappa ' + \dot{\Phi}_{20'}{} \bar{\kappa}' + \thop \
\dot{\Phi}_{21'}{} + \thop \dot{\Psi}_{3}{} -  \eth \dot{\Psi}_{4}{} \
-  \eth' \dot{\Phi}_{22'}{}\, ,
\end{equation}
which allows the elimination of explicit dependence on the perturbed GHP derivatives $\dot{\eth}$ (introduced in \cite{Fransen:2025cgv} to preserve GHP covariance under perturbations). We further used that, on \eqref{eq:PLLR}, $\Psi_{4}$ is constant and $\eth'\bar{\kappa}'=0$.

The left-hand sides of the above equations involve quantities that are generally nonzero for vacuum GR perturbations, whereas the right-hand sides collected into the source terms $S_{0},\,S_{1},\,\ldots$ are proportional to variations of the Ricci tensor and therefore vanish in vacuum GR. Note that the left-hand sides involve not only Weyl scalars but also the quantities $\dot{\tau}$ and $\dot{\sigma}$. Equations for these follow from linearized Ricci identities,
\begin{equation}\label{eq:thosigmatau}
    \tho \dot{\sigma}=\dot{\Psi}_{0}\, ,\qquad \tho\dot{\tau}=\dot{\Psi}_{1}-\dot{\Phi}_{01'}\, .
\end{equation}
We emphasize that Eqs.~\eqref{eq:Psi0CWE}-\eqref{eq:Psi4CWE} and \eqref{eq:thosigmatau} are identities valid for any perturbation of the background geometry \eqref{eq:PLLR} in the GPT gauge, without assuming equations of motion. This formulation is particularly convenient because the system can be solved iteratively: assuming knowledge of all Ricci tensor perturbations, one first solves for $\dot{\Psi}_{0}$, then uses that solution to determine $\dot{\Psi}_{1}$, followed by $\dot{\Psi}_{2}$. One then solves \label{eq:sigmatau} for $\dot{\sigma}$ and $\dot{\tau}$, which in turn allows determination of $\dot{\Psi}_{3}$ and finally $\dot{\Psi}_{4}$. This nested structure is sufficient for our purposes.

We now specialize to perturbations satisfying \eqref{eq:HD_Eqs}. The perturbations are decomposed as
\begin{equation}\label{eq:pertsnonvac}
    \dot{\Psi}_{n}=\dot{\Psi}^\text{\tiny GR}_{n}+\ell_\text{\tiny UV}^{6}\dot{\Psi}^\text{\tiny HD}_{n}\, ,\qquad \dot{\sigma}=\dot{\sigma}^\text{\tiny GR}+\ell_\text{\tiny UV}^{6}\dot{\sigma}^\text{\tiny HD}\, ,\qquad \dot{\tau}=\dot{\tau}^\text{\tiny GR}+\ell_\text{\tiny UV}^{6}\dot{\tau}^\text{\tiny HD}\, ,
\end{equation}
with $n=0,\ldots,4$, where the superscripts $\text{\tiny GR}$ and $\text{\tiny HD}$ denote the GR solution and the higher-derivative correction, respectively. We have already set to zero the cubic corrections $\sim\ell_\text{\tiny UV}^{4}$, since, as discussed in the main text, they do not affect the dynamics. The Ricci tensor is expanded as
\begin{equation}
    \dot{R}_{ab}= \ell_\text{\tiny UV}^{6}\left(\dot{P}^{(8)}_{ab}\left[h^\text{\tiny GR}\right]-\frac{1}{2}\tensor{\dot{P}}{^{(8)c}_c}\left[h^\text{\tiny GR}\right]g_{ab}\right)\, .
\end{equation}
With this expansion, the perturbations $\dot{\Phi}_{00'}, \ldots$ can be expressed entirely in terms of the GR solution. Therefore, once $h^\text{\tiny GR}$ is known, all Ricci perturbations are determined up to order $\ell_\text{\tiny UV}^{6}$. The higher-derivative contributions in Eq.~\eqref{eq:pertsnonvac} then satisfy Eqs.~\eqref{eq:Psi0CWE}--\eqref{eq:Psi4CWE} and \eqref{eq:thosigmatau}, with source terms fully determined by the GR solution, which itself is constructed from the Hertz potential. Consequently, all $S_{0}, S_{1}, \ldots$ are known in terms of the Hertz potential, and the higher-derivative corrections can be obtained by solving the system iteratively as described above.

For concreteness, we provide the explicit source terms, evaluated for Hertz potentials satisfying $\ell^{a}\nabla_{a}\Psi_\text{\tiny H}=i p_{v}\Psi_\text{\tiny H}$, as in the QNM ansatz \eqref{eq:Hertz_Sep} (in fact, since $[\ell^{a}\nabla_{a},\nabla_{b}]=0$ because $\ell^{a}R_{abcd}=0$ and $\nabla_{a}\ell_{b}=0$, this assumption does not restrict generality):
\begin{align}\label{eq:source0}
    S_{0}&=32 p_{v}^{8}\Omega^{4}\left(\epsilon_{1}+\epsilon_{2}\right)\bar{\Psi}_\text{\tiny H}+32 p_{v}^{8}\Omega^{4}\left(\epsilon_{1}-\epsilon_{2}\right)\Psi_\text{\tiny H}\, ,\\ \notag \\
    S_{1}&=32 i p_{v}^{7}\Omega^{4}\left(\epsilon_{1}+\epsilon_{2}\right)\bar{m}^{a}\nabla_{a}\bar{\Psi}_\text{\tiny H}-32 i p_{v}^{7}\Omega^{4}\left(\epsilon_{1}-\epsilon_{2}\right)\bar{m}^{a}\nabla_{a}\Psi_\text{\tiny H}\, ,\\  \notag \\
    S_{2}&=-32 p_{v}^{6}\Omega^{4}\left(\epsilon_{1}+\epsilon_{2}\right)\bar{m}^{a}\bar{m}^{b}\nabla_{a}\nabla_{b}\bar{\Psi}_\text{\tiny H}-32 p_{v}^{6}\Omega^{4}\left(\epsilon_{1}-\epsilon_{2}\right)\bar{m}^{a}\bar{m}^{b}\nabla_{a}\nabla_{b}\Psi_\text{\tiny H}\, ,\\ \notag \\
    S_{3}&=-32 i p_{v}^{5}\Omega^{4}\left(\epsilon_{1}+\epsilon_{2}\right)\bar{m}^{a}\bar{m}^{b}\bar{m}^{c}\nabla_{a}\nabla_{b}\nabla_{c}\bar{\Psi}_\text{\tiny H}+32 i p_{v}^{5}\Omega^{4}\left(\epsilon_{1}-\epsilon_{2}\right)\bar{m}^{a}\bar{m}^{b}\bar{m}^{c}\nabla_{a}\nabla_{b}\nabla_{c}\Psi_\text{\tiny H}\, ,\\  \notag \\
    S_{4}&=0\, .   
\end{align}
To complete the specification of the source terms, one also requires $\dot{\Phi}_{01'}$, which appears in the equation for $\dot{\tau}$ in Eq.~\eqref{eq:thosigmatau}. This component vanishes, $\dot{\Phi}_{01}'=0$. We notice that, although $S_{4}=0$, the quantity $\dot{\Psi}^\text{\tiny HD}_{4}$ still receives corrections through the terms $\dot{\Psi}^\text{\tiny HD}_{2}$, $\dot{\Psi}^\text{\tiny HD}_{3}$, and $\dot{\tau}^\text{\tiny HD}$ appearing on the left-hand side of Eq.~\eqref{eq:Psi4CWE}, which are themselves obtained by solving the nested system.

\section{Duality transformation of $\dot{C}^{\pm}$ in GR}\label{app:dualityCs}
\noindent
The general expression for the self-dual Weyl tensor in terms of a null frame and the associated Weyl scalars is 
\begin{equation}\label{eq:Cplus}
        C^{+}_{abcd}=-\bar{\Psi}_{0}\bar{Z}^{-}_{ab}\bar{Z}^{-}_{cd}-\bar{\Psi}_{1}\left(\bar{Z}^{-}_{ab}\bar{Z}^{\circ}_{cd}+\bar{Z}^{\circ}_{ab}\bar{Z}^{-}_{cd}\right)-\bar{\Psi}_{2}\left(\bar{Z}^{\circ}_{ab}\bar{Z}^{\circ}_{cd}+\bar{Z}^{+}_{ab}\bar{Z}^{-}_{cd}+\bar{Z}^{-}_{ab}\bar{Z}^{+}_{cd}\right)-\bar{\Psi}_{3}\left(\bar{Z}^{+}_{ab}\bar{Z}^{\circ}_{cd}+\bar{Z}^{\circ}_{ab}\bar{Z}^{+}_{cd}\right)-\bar{\Psi}_{4}\bar{Z}^{+}_{ab}\bar{Z}^{+}_{cd}\, ,
\end{equation}
where $Z^{+}_{ab}=(\ell\wedge m)_{ab}$, $Z^{-}_{ab}=(\bar{m}\wedge n)_{ab}$, and $Z^{\circ}_{ab}=(m\wedge \bar{m})_{ab}-(\ell\wedge n)_{ab}$ form a basis of self-dual bivectors with GHP weights $(2,0)$, $(-2,0)$, and $(0,0)$, respectively. As usual, the overbar denotes complex conjugation, which acts on GHP weights as $(p,q)\mapsto(q,p)$.

We now compute the perturbation of Eq.~\eqref{eq:Cplus} induced by a general gravitational perturbation around the background~\eqref{eq:PL}, without assuming any specific equations of motion. In the GPT gauge introduced above, which is purely geometric and does not rely on the dynamics, one has $\dot{\ell}^{a}=\left(\ell_{a}\right)\dot{}=0$. Using the fact that the background Weyl scalars $\Psi_{0,1,2,3}$ vanish, one obtains the following expression for $\dot{C}^{+}$:
\begin{equation}
    \begin{aligned}\label{eq:delta_Cplus}
        \dot{C}^{+}_{abcd}&=-\dot{\bar{\Psi}}_{0}\bar{Z}^{-}_{ab}\bar{Z}^{-}_{cd}-\dot{\bar{\Psi}}_{1}\left(\bar{Z}^{-}_{ab}\bar{Z}^{\circ}_{cd}+\bar{Z}^{\circ}_{ab}\bar{Z}^{-}_{cd}\right)-\dot{\bar{\Psi}}_{2}\left(\bar{Z}^{\circ}_{ab}\bar{Z}^{\circ}_{cd}+\bar{Z}^{+}_{ab}\bar{Z}^{-}_{cd}+\bar{Z}^{-}_{ab}\bar{Z}^{+}_{cd}\right)-\dot{\bar{\Psi}}_{3}\left(\bar{Z}^{+}_{ab}\bar{Z}^{\circ}_{cd}+\bar{Z}^{\circ}_{ab}\bar{Z}^{+}_{cd}\right)\\
        &-\dot{\bar{\Psi}}_{4}\bar{Z}^{+}_{ab}\bar{Z}^{+}_{cd}-\bar{\Psi}_{4}\left[(\ell\wedge (\bar{m})\dot{})_{ab}\bar{Z}^{+}_{cd}+\bar{Z}^{+}_{ab}(\ell\wedge (\bar{m})\dot{})_{cd}\right]\, .
    \end{aligned}
\end{equation}
This expression is completely general and can be evaluated once specific equations of motion are imposed. In the case of GR, the Weyl scalars are given in \eqref{eq:Weylperts}, where all of them are proportional to derivatives of $\bar{\Psi}_\text{\tiny H}$, and $\Psi_\text{\tiny H}$ does not appear. Consequently, their complex conjugates appearing in \eqref{eq:delta_Cplus} are proportional to derivatives of $\Psi_\text{\tiny H}$ only. 

Furthermore, in the GPT gauge one has $\left(\bar{m}_{a}\right)\dot{}=-\dot{\bar{m}}_{a}$, with $\dot{\bar{m}}_{a}$ given in \eqref{frame2}. Evaluating this expression on the metric perturbation~\eqref{eq:recmet2_final} yields~\cite{Fransen:2025cgv}
\begin{equation}
    \dot{\bar{m}}_{a}=\frac{1}{2}\left(\ell_{a}\tho\eth \Psi_\text{\tiny H}-m_{a}\tho^{2}\Psi_\text{\tiny H}\right)\, ,
\end{equation}
which only contains $\Psi_\text{\tiny H}$ and not $\bar{\Psi}_\text{\tiny H}$. Substituting this result into Eq.~\eqref{eq:delta_Cplus}, one finds that $\dot{C}^{+}_{abcd}$ depends solely on derivatives of $\Psi_\text{\tiny H}$, with no contribution from $\bar{\Psi}_\text{\tiny H}$. Since for the anti-self-dual component one has $\dot{C}^{-}_{abcd}=\dot{\bar{C}}^{+}_{abcd}$, the complementary statement holds: $\dot{C}^{-}_{abcd}$ depends only on $\bar{\Psi}_\text{\tiny H}$. 

This establishes that, under~\eqref{eq:EMDualityHertz_final}, the perturbations transform according to the gravitational duality
\begin{equation}
    \dot{C}^{\pm}_{abcd}\mapsto e^{\pm i \theta } \dot{C}^{\pm}_{abcd}\, .
\end{equation}
As we show below, the same argument extends to the case where higher-derivative corrections are included.

\section{Duality transformation of $\dot{C}^{\pm}$ beyond GR}\label{app:dualityCs_bGR}
\noindent
We now use the nested system of equations introduced above to show that the symmetry~\eqref{eq:EMDualityHertz_final} induces a duality rotation of the higher-derivative correction to the (anti)self-dual Weyl tensor, $\dot{C}^{\pm \text{\tiny HD}}$, only if $\epsilon_{1}=\epsilon_{2}$. 

We begin by observing that, only in the case $\epsilon_{1}=\epsilon_{2}$, the source term~\eqref{eq:source0} for $\dot{\Psi}^{\text{\tiny HD}}_{0}$ depends exclusively on $\bar{\Psi}_\text{\tiny H}$ and not on $\Psi_\text{\tiny H}$. Consequently, only under this condition does Eq.~\eqref{eq:EMDualityHertz_final} imply $\dot{\Psi}^{\text{\tiny HD}}_{0}\mapsto e^{-i\theta}\dot{\Psi}^{\text{\tiny HD}}_{0}$. This property propagates through the nested system of equations, leading to the conclusion that, again only if $\epsilon_{1}=\epsilon_{2}$, one has $\dot{\Psi}^{\text{\tiny HD}}_{n}\mapsto e^{-i\theta}\dot{\Psi}^{\text{\tiny HD}}_{n}$ for all $n=0,\ldots,4$. 

However, inspection of Eq.~\eqref{eq:delta_Cplus} shows that this is not yet sufficient to ensure that $\dot{C}^{+\text{\tiny HD}}$ transforms as a duality rotation under~\eqref{eq:EMDualityHertz_final}. It is therefore necessary to determine how the higher-derivative correction to $(\bar{m}_{a})\dot{}$ transforms. Since in the GPT gauge $(\bar{m}_{a})\dot{}=-\dot{\bar{m}}_{a}$, we focus on $\dot{\bar{m}}_{a}$, which is given by Eq.~\eqref{frame2},
\begin{equation}
    \dot{\bar{m}}^{a}=\frac{1}{2}\left(-h_{n\bar{m}}\ell^{a}+h_{m\bar{m}}\bar{m}^{a}+h_{\bar{m}\bar{m}}m^{a}\right)\, .
\end{equation}
To proceed, we linearize the equation expressing $\nabla_{a}m_{b}$ in terms of GHP quantities,
\begin{equation}
\begin{aligned}
    \nabla_{a}\bar{m}_{b}&=\ell_{a} (- \bar{m}_{b} n^{c} \
\omega_{c} + \bar{m}_{b} n^{c} \bar{\omega}_{c} -  \ell_{b} \kappa ' -  \
n_{b} \bar{\tau})   + n_{a} (- \
\ell^{c} \bar{m}_{b} \omega_{c} + \ell^{c} \bar{m}_{b} \bar{\omega}_{c} -  \
n_{b} \bar{\kappa} -  \ell_{b} \tau ') \\
&+ m_{a} \
(\bar{m}_{b} \bar{m}^{c} \omega_{c} -  \bar{m}_{b} \bar{m}^{c} \
\bar{\omega}_{c} + \ell_{b} \sigma ' + n_{b} \bar{\sigma}) + \bar{m}_{a} (m^{c} \bar{m}_{b} \omega_{c} -  m^{c} \bar{m}_{b} \
\bar{\omega}_{c} + \ell_{b} \rho ' + n_{b} \bar{\rho})  
\end{aligned}
\end{equation}
where $\omega_{a}$ is the GHP connection 1-form. Linearizing this relation, projecting onto $(\bar{m}^{a}\ell^{b}-\ell^{a}\bar{m}^{b})$ and $(m^{a}\ell^{b}-\ell^{a}m^{b})$, and using the GPT gauge conditions \eqref{eq:geod_Trans} and \eqref{eq:zeroSC}, one obtains the following equations for $h_{\bar{m}\bar{m}}$ and $h_{m\bar{m}}$:
\begin{equation}\label{eq:hmm_eqs}
    \tho h_{\bar{m}\bar{m}}=2\dot{\bar{\sigma}}\, ,\qquad \tho h_{m\bar{m}}=2\dot{\bar{\rho}}\, .
\end{equation}
These relations hold for arbitrary perturbations of the metric~\eqref{eq:PLLR} in the GPT gauge. An equation for $\dot{\sigma}$ is already given in~\eqref{eq:thosigmatau}, while an equation for $\dot{\rho}$ follows directly from the linearization of a Ricci identity,
\begin{equation}
    \tho\dot{\rho}=-\dot{\Phi}_{00'}\, .
\end{equation}
We can now complete the proof that $\dot{C}^{+\text{\tiny HD}}\mapsto e^{i\theta}\dot{C}^{+\text{\tiny HD}}$. First, we decompose the relevant quantities as
\begin{equation}
    h_{\bar{m}\bar{m}}= h^{\text{\tiny GR}}_{\bar{m}\bar{m}}+\ell^{6}_{\text{\tiny UV}} h^{\text{\tiny HD}}_{\bar{m}\bar{m}}\,, \quad h_{m\bar{m}}= h^{\text{\tiny GR}}_{m\bar{m}}+\ell^{6}_{\text{\tiny UV}} h^{\text{\tiny HD}}_{m\bar{m}}\, ,\quad \dot{\sigma}=\dot{\sigma}^{\text{\tiny GR}}+\ell^{6}_{\text{\tiny UV}}\dot{\sigma}^{\text{\tiny HD}}\, \quad \dot{\rho}=\dot{\rho}^{\text{\tiny GR}}+\ell^{6}_{\text{\tiny UV}}\dot{\rho}^{\text{\tiny HD}}\, .
\end{equation}
Using the fact that $\dot{\Phi}_{00'}=0$ (as computed above), it follows that $\dot{\rho}$ receives no higher-derivative correction, so one can set $\dot{\rho}^{\text{\tiny HD}}=0$. Then, from Eq.~\eqref{eq:hmm_eqs}, one can also set $h^{\text{\tiny HD}}_{m\bar{m}}=0$. 

On the other hand, as previously established, when $\epsilon_{1}=\epsilon_{2}$ one has $\dot{\Psi}^{\text{\tiny HD}}_{0}\mapsto e^{-i\theta }\dot{\Psi}^{\text{\tiny HD}}_{0}$ under \eqref{eq:EMDualityHertz_final}. By Eq.~\eqref{eq:thosigmatau}, this implies $\dot{\sigma}^{\text{\tiny HD}}\mapsto e^{-i\theta }\dot{\sigma}^{\text{\tiny HD}}$. Substituting into Eq.~\eqref{eq:hmm_eqs}, we deduce that
\begin{equation}
    h^{\text{\tiny HD}}_{\bar{m}\bar{m}}\mapsto e^{i\theta}h^{\text{\tiny HD}}_{\bar{m}\bar{m}}\,.
\end{equation}
With this result, the higher-derivative contribution to $\dot{C}^{+}_{abcd}$, as given by Eq.~\eqref{eq:delta_Cplus}, takes the form
\begin{equation}
    \begin{aligned}\label{eq:delta_Cplus_HD}
        \dot{C}^{+\text{\tiny HD}}_{abcd}&=-\dot{\bar{\Psi}}^{\text{\tiny HD}}_{0}\bar{Z}^{-}_{ab}\bar{Z}^{-}_{cd}-\dot{\bar{\Psi}}^{\text{\tiny HD}}_{1}\left(\bar{Z}^{-}_{ab}\bar{Z}^{\circ}_{cd}+\bar{Z}^{\circ}_{ab}\bar{Z}^{-}_{cd}\right)-\dot{\bar{\Psi}}^{\text{\tiny HD}}_{2}\left(\bar{Z}^{\circ}_{ab}\bar{Z}^{\circ}_{cd}+\bar{Z}^{+}_{ab}\bar{Z}^{-}_{cd}+\bar{Z}^{-}_{ab}\bar{Z}^{+}_{cd}\right)-\dot{\bar{\Psi}}^{\text{\tiny HD}}_{3}\left(\bar{Z}^{+}_{ab}\bar{Z}^{\circ}_{cd}+\bar{Z}^{\circ}_{ab}\bar{Z}^{+}_{cd}\right)\\
        &-\dot{\bar{\Psi}}^{\text{\tiny HD}}_{4}\bar{Z}^{+}_{ab}\bar{Z}^{+}_{cd}+\frac{h^{\text{\tiny HD}}_{\bar{m}\bar{m}}}{2}\bar{\Psi}_{4}\left(Z^{+}_{ab}\bar{Z}^{+}_{cd}+\bar{Z}^{+}_{ab}Z^{+}_{cd}\right)\, ,
    \end{aligned}
\end{equation}
from which it is manifest that \eqref{eq:EMDualityHertz_final} induces the transformation $\dot{C}^{\pm \text{\tiny HD}}_{abcd}\mapsto e^{\pm i\theta}\dot{C}^{\pm \text{\tiny HD}}_{abcd}$.

\section{Shift to the QNM frequency}\label{app:Frequency_shift}
\noindent
It is sufficient to consider the equation for $\dot{\Psi}^{\text{\tiny HD}}_{0}$ in \eqref{eq:Psi0CWE}, with the source \eqref{eq:source0}, already imposing $\epsilon_{1}=\epsilon_{2}\equiv\epsilon$:
\begin{equation}\label{eq:Psi0QNMHD}
    \square \dot{\Psi}^{\text{\tiny HD}}_{0}=\tilde{\epsilon} \ 64 p_{v}^{8}\Omega^{4}\bar{\Psi}_\text{\tiny H}\,,
\end{equation}
where $\tilde{\epsilon}=\epsilon/\ell^{6}_{\text{\tiny UV}}$ is the dimensionless coupling. We assume that $\bar{\Psi}_\text{\tiny H}$ corresponds to a QNM solution~\eqref{eq:Hertz_Sep} with frequencies~\eqref{eq:onshell_freq} (see e.g.~\cite{Fransen:2025cgv} for the explicit solution). Using $\tho \Psi_\text{\tiny H}=i p_{v}\Psi_\text{\tiny H}$ and Eq.~\eqref{eq:Weylperts}, Eq.~\eqref{eq:Psi0QNMHD} can be written as
\begin{equation}\label{eq:correctedQNM_EQ}
    \square \dot{\Psi}^{\text{\tiny HD}}_{0}=-\tilde{\epsilon} \ 128 p_{v}^{4}\Omega^{4}\dot{\Psi}^{\text{\tiny GR}}_{0}\,.
\end{equation}
The GR solution for $\dot{\Psi}_{0}$ has the form 
\begin{equation}
     \dot{\Psi}^{\text{\tiny GR}}_{0}=e^{-i p^\text{\tiny GR}_{u}u-i p_{v}v}X_{n_{x}}(x)R^{\text{\tiny GR}}(y)\, ,
\end{equation}
where $p^{\text{\tiny GR}}_{u}$ denotes the GR value of the frequency~\eqref{eq:onshell_freq}. The form of the QNM correction then follows from the expansion, in a manner similar in spirit to \cite{Zimmerman:2014aha},
\begin{equation}
\begin{aligned}
    \dot{\Psi}_{0}&=e^{-i p_{u}u-i p_{v}v}X_{n_{x}}(x)R(y)=e^{-i (p^{\text{\tiny GR}}_{u}+\ell_{\text{\tiny UV}}^{6}\alpha )u-i p_{v}v}X_{n_{x}}(x)(R^{\text{\tiny GR}}(y)+\ell_{\text{\tiny UV}}^{6}R^{\text{\tiny HD}}(y))\\
    &=\dot{\Psi}^{\text{\tiny GR}}_{0}+\ell^{6}_{\text{\tiny UV}}\left(-i\alpha u \ \dot{\Psi}^{\text{\tiny GR}}_{0}+e^{-i p^{\text{\tiny GR}}_{u}u-i p_{v}v}X_{n_{x}}(x)R^{\text{\tiny HD}}(y)\right)+\text{higher orders in $\ell_{\text{\tiny UV}}$}\,,
\end{aligned}
\end{equation}
so the ansatz for $\dot{\Psi}^{\text{\tiny HD}}_{0}$ is
\begin{equation}\label{eq:ansatz}
   \dot{\Psi}^{\text{\tiny HD}}_{0} =-i\alpha u \ \dot{\Psi}^{\text{\tiny GR}}_{0}+e^{-i p^{\text{\tiny GR}}_{u}u-i p_{v}v}X_{n_{x}}(x)R^{\text{\tiny HD}}(y)\, ,
\end{equation}
where the parameter $\alpha$ determines the frequency shift through $\delta p_{u}=\ell^{6}_{\text{\tiny UV}}\alpha$, and $R^{\text{\tiny HD}}(y)$ is the correction to the wavefunction. The first term in Eq.~\eqref{eq:ansatz} has the form of a resonant solution, with a contribution linear in $u$. This is expected for a frequency shift, since the source term on the right-hand side of Eq.~\eqref{eq:correctedQNM_EQ} has a frequency that coincides with a pole of the propagator on the left-hand side (i.e. a resonance). Substituting Eq.~\eqref{eq:ansatz} into Eq.~\eqref{eq:correctedQNM_EQ}, one finds that there is no correction to the wavefunction, $R^{\text{\tiny HD}}(y)=0$, and that the frequency shift is
\begin{equation}
    \delta p_{u}= 64 \Omega^{4} p_{v}^{3} \ \epsilon\, .
\end{equation}

\end{document}